\documentclass[preprint,review,12pt]{elsarticle}
\pdfoutput=1

\usepackage{graphicx}

\usepackage{multirow}

\usepackage{amssymb}

\begin{document}

\begin{frontmatter}


\title{Influence of solvation on the structural and capacitive properties of electrical double layer capacitors}
\author[ISE1,PECSA,RS2E,cor1]{C\'{e}line Merlet\corref{cor1}}
\ead{celine.merlet@upmc.fr}
\cortext[cor1]{Corresponding author}
\author[ISE1,PECSA,RS2E]{Mathieu Salanne}
\author[PECSA,RS2E]{Benjamin Rotenberg}
\author{Paul A. Madden\fnref{MML}}
\fntext[ISE1]{ISE member}
\fntext[PECSA]{UPMC Univ Paris 06 and CNRS, UMR 7195, PECSA, F-75005, Paris, France}
\fntext[RS2E]{R\'eseau sur le Stockage Electrochimique de l'Energie (RS2E), FR CNRS 3459, France}
\fntext[MML]{Department of Materials, University of Oxford, Parks Road, Oxford OX1 3PH, UK}
\fntext[cor1]{Laboratoire PECSA - CC51, Universit\'e Pierre et Marie Curie, 75005 Paris, France\\ (tel) +33 1 44 27 31 17, (fax) +33 1 44 27 32 28}

\begin{abstract}
We use molecular dynamics simulations to explore the impact of a non-ionic solvent on the structural and capacitive properties of supercapacitors based on an ionic liquid electrolyte and carbon electrodes. The study is focused on two pure ionic liquids, namely 1-butyl-3-methylimidazolium hexafluorophosphate and 1-butyl-3-methylimidazolium tetrafluoroborate, and their 1.5~M solutions in acetonitrile. The electrolytes, represented by coarse-grained models, are enclosed between graphite electrodes. We employ a constant potential methodology which allows us to gain insight into the influence of solvation on the polarization of the electrodes as well as the structural and capacitive properties of the electrolytes at the interface. We show that the interfacial characteristics, different for two distinct pure ionic liquids, become very similar upon mixing with acetonitrile. 
\end{abstract}

\begin{keyword}
double layer capacitance \sep solvation  \sep molecular dynamics simulations \sep interfacial structure \sep graphite
\end{keyword}

\end{frontmatter}


\newpage

\section{Introduction}
\label{Intro}

Supercapacitors store energy at the electrode/electrolyte interface without involving faradaic reactions; this confers on them characteristics very distinct from batteries. They can be used as high-power generators and can undergo one million charge/discharge cycles without deterioration. Nevertheless, compared to batteries, they suffer from a relatively low energy density. There are four principal components in supercapacitors on which we can act to optimize these systems: the active matter, the current collectors, the separator and the electrolyte. As the energy stored in a supercapacitor is proportional to the capacitance and the square of the operating voltage ($E=\frac{1}{2}CU^2$), the improvements will come by optimizing the electrode morphology, which determines the capacity of the system, and the electrolyte, which sets the maximum voltage by its decomposition limit~\cite{Simon12}. The modifications of these two components will also impact the power which can be delivered by the capacitor, as this property is a function of the maximum voltage and resistance ($P=\frac{U}{4R}$).

Focusing on liquid electrolytes, different fluids have been studied up to date which are suited for distinct applications. Aqueous electrolytes are attractive because of their high ionic conductivities ($>$~400~mS.cm$^{-1}$), which allow for a higher specific power, but they have relatively narrow electrochemical windows (1.2~V)~\cite{Wang12b}. On the contrary, ionic liquids (ILs) and organic electrolytes exhibit larger electrochemical windows, up to 5~V and 3~V respectively~\cite{Galinski06,Zhang09,Balducci07}. Ionic liquids have a number of attractive properties such as low combustibility, high thermal stability and low vapor pressure, which make them \emph{a priori} safe. They are also adaptable thanks to the broad choice of anions and cations that can be combined. From mixtures of different ILs, various operating conditions can be ameliorated, for example the working temperature range can be enlarged~\cite{Lin11}. The major drawback of using ILs in supercapacitors is their low ionic conductivity ($<$~15~mS.cm$^{-1}$)~\cite{Galinski06}.

Consequently, in many experimental studies and applications of supercapacitors, organic electrolytes using acetonitrile (ACN) or propylene carbonate (PC) as solvent are still used instead of pure ILs~\cite{Chmiola06,Centeno11}.  It has been found that adding ACN to ILs enhances greatly the ionic conductivity of the system~\cite{Chaban12,Wang03}. The structural and dynamic effects of solvation on the bulk properties of electrolytes have been examined by experiments~\cite{Wang03,Huo07,Sadeghi11} and simulations~\cite{Chaban12,Wu05,Chaban11b} but the interfacial properties of the mixtures are less thoroughly studied, especially from a theoretical standpoint. Recent electrochemical measurements on carbide-derived carbon electrodes immersed in pure 1-ethyl-3-methylimidazolium bis(trifluoromethylsulfonyl)imide ([EMI][TFSI]) and in a 2~M solution of [EMI][TFSI] in ACN display an unexpected peak in the cyclic voltammetry (CV) curve of the solution~\cite{Lin09b}. Some experimental results indicates that this peak is not associated with a faradaic process and a clear explanation is missing. Molecular simulations of the interface between porous carbons and organic electrolytes may be needed to explain this effect, but, even simpler interfaces between, for example, planar electrodes and solutions may contribute to a better understanding.

The Gouy-Chapman theory, which describes highly-diluted solutions at smooth planar interfaces, is not applicable for common electrolytes as typical salt concentrations are 1~M or more. To study these interfaces, there is a need for more complex theories~\cite{Kornyshev07} or molecular simulations which correctly describe ionic correlations. Up-to-date, molecular simulations involving the interface between various carbon structures (carbon nanotubes~\cite{Frolov12,Yang09}, slit pore~\cite{Jiang12b,Tanaka10}, graphite~\cite{Feng10b,Feng11b,Shim11,Shim12}) and organic electrolytes have mainly focused on the liquid side of the interface. In the present study, we investigate the influence of mixing two salts, namely 1-butyl-3-methylimidazolium hexafluorophosphate ([BMI][PF$_6$]) and 1-butyl-3-methylimidazolium tetrafluoroborate ([BMI][BF$_4$]), with ACN on the electrolyte and electrode properties. We use a comparison between pure ILs and their corresponding 1.5~M organic electrolytes to highlight the modifications of structural and capacitive properties associated with solvation. The use of a constant potential approach to model the electrodes, instead of the constant charge methodology, allows us to gain knowledge about the polarization of the electrodes by the electrolytes and how this is affected by the ACN solvent. 

\section{Computational details}
\label{Comput}

\begin{figure*}
\begin{center}
\includegraphics[scale=0.42]{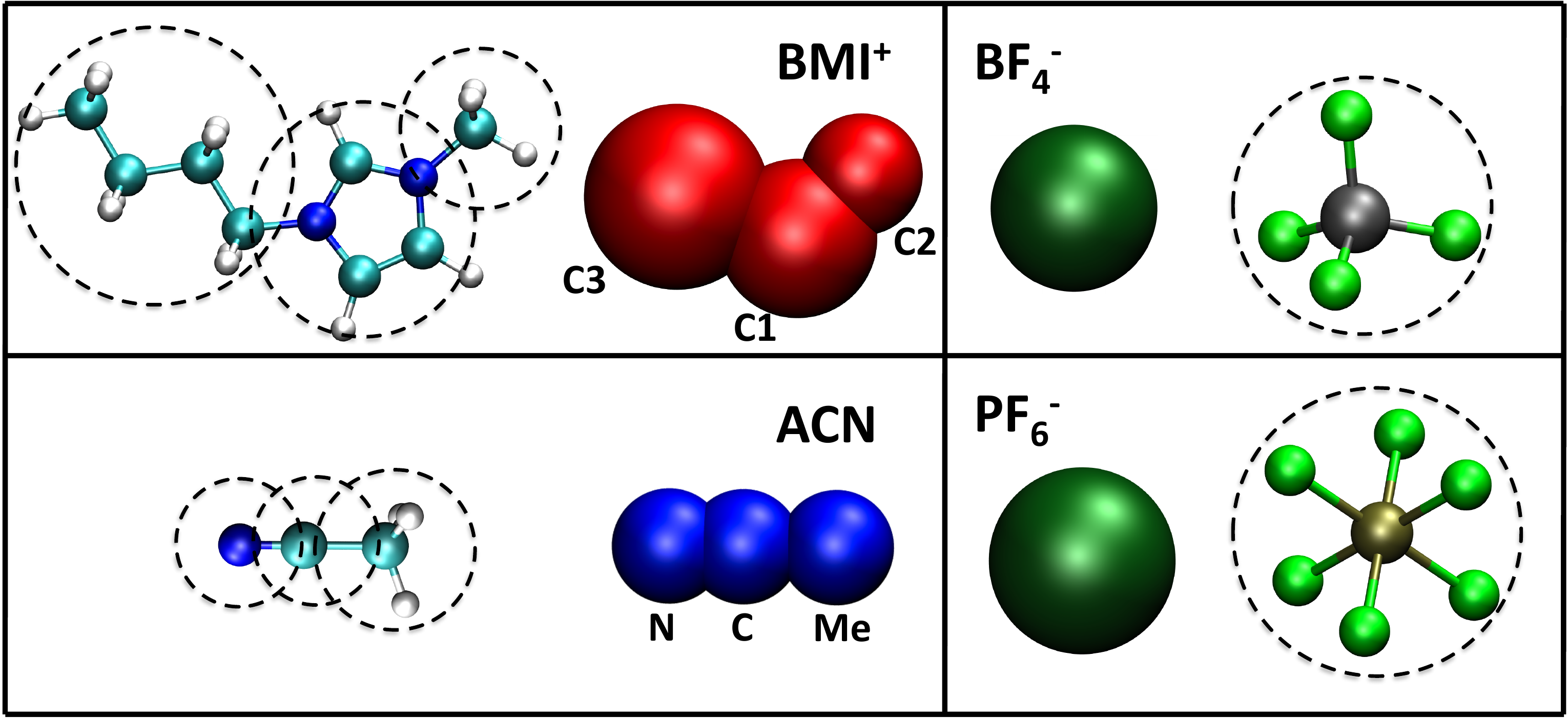}
\end{center}
\caption{Schematic representations of the coarse-grained models for the molecules studied in this work.}
\label{CGM}
\end{figure*}

Molecular dynamics simulations are conducted on four different electrolytes surrounded by model graphite electrodes: pure [BMI][PF$_6$] and [BMI][BF$_4$] and their corresponding 1.5~M solutions with ACN as a solvent. All molecules are represented by a coarse-grained model in which the forces are calculated as the sum of site-site Lennard-Jones potential and coulombic interactions. Parameters for the ions and carbon atoms are the same as in our previous works~\cite{Merlet11,Merlet12b}: three sites are used to describe the ACN and the cation and the anions are treated as spheres. The two ILs differ by the nature of the anion, the diameter of PF$_6^-$ being larger than the one of BF$_4^-$ by approximately 0.5~\r{A}. The model for ACN was developed by Edwards \emph{et al.}~\cite{Edwards84}. All the parameters of the force field are recalled in table~\ref{param} and a schematic representation of the molecules is given in figure~\ref{CGM}.

\begin{figure*}
\begin{center}
\includegraphics[scale=0.42]{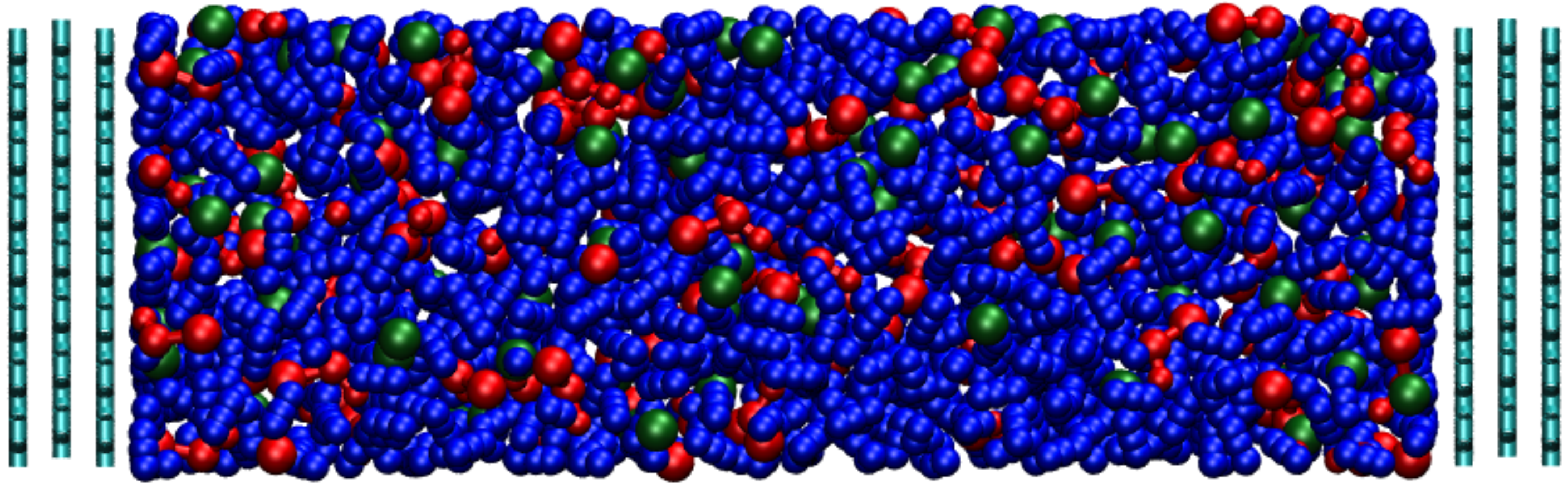}
\end{center}
\caption{Snapshot of the simulation cell: 1.5~M solution of BMI$^+$ cations (red) and BF$_4^-$ anions (green) in ACN (dark blue) enclosed between graphite walls (light blue).}
\label{snapshot}
\end{figure*}

Each electrode is modelled as three fixed graphene layers. The electrolyte is enclosed between two planar electrodes and two-dimensional periodic boundary conditions are applied, i.e. there is no periodicity in the direction perpendicular to the electrodes. Figure~\ref{snapshot} shows a snapshot of the simulation cell for the ACN-[BMI][BF$_4$] mixture with a salt concentration of 1.5~M. The molecular dynamics simulations were conducted in the NVT ensemble using a time step of 2~fs and a Nos\'e-Hoover thermostat~\cite{Martyna92} with a time constant of 10~ps. The Ewald summation is done consistently with the two-dimensional periodic boundary conditions~\cite{Reed07,Gingrich10}.

Pure ILs and electrolyte solutions are simulated at 400~K and 298~K respectively. These temperatures were chosen because of the very high viscosities of the pure ILs at room temperature (261.4~mPa.s for [BMI][PF$_6$] and 100.2~mPa.s for [BMI][BF$_4$]~\cite{Tokuda04}) and the fact that ACN boils at 355~K. Nevertheless, static properties usually depend weakly on the temperature for conditions far from phase transitions. Thus, the qualitative conclusions raised in this article should hold for pure ILs  at lower temperatures and the comparison with ACN-based electrolytes is relevant. The sizes of the simulation cells are chosen in order to reproduce the experimental densities of the electrolytes. Table~\ref{cells} gathers the lengths and number of molecules for all the simulation cells.

\begin{table*}
\begin{center}
\begin{tabular}{|c|c|c|c|c|c|c|c|c|}
\hline
Site & C1 & C2 & C3 & PF$_6^-$ & BF$_4^-$ & N & C & Me \\
\hline
q (e) & 0.4374 & 0.1578 & 0.1848 & -0.78 & -0.78 & -0.398 & 0.129 & 0.269 \\
\hline
M (g.mol$^{-1}$) & 67.07 & 15.04 & 57.12 & 144.96 & 86.81 & 14.01 & 12.01 & 15.04 \\
\hline
$\sigma_i$ (\r{A}) & 4.38 & 3.41 & 5.04 & 5.06 & 4.51 & 3.30 & 3.40 & 3.60 \\  
\hline
$\varepsilon_i$ (kJ.mol$^{-1}$) & 2.56 & 0.36 & 1.83 & 4.71 & 3.24 & 0.42 & 0.42 & 1.59 \\
\hline
\end{tabular}
\end{center}
\caption{Force-field parameters for the molecules of the electrolytes~\cite{Merlet11,Edwards84,Roy10b} (geometries of the molecules are available in the aforementioned publications). Site-site interaction energies are given by the sum of a Lennard-Jones potential and coulombic interactions $u_{ij}(r_{ij})=4\varepsilon_{ij}[(\frac{\sigma_{ij}}{r_{ij}})^{12}-(\frac{\sigma_{ij}}{r_{ij}})^6]+\frac{q_iq_j}{4\pi\varepsilon_0r_{ij}}$ where $r_{ij}$ is the distance between sites, $\varepsilon_0$ is the permittivity of free space and crossed parameters are calculated by Lorentz-Berthelot mixing rules. The parameters for the carbon atoms of the graphite electrodes are $\sigma_{\rm C}$ = 3.37~\r{A} and $\varepsilon_{\rm C}$ = 0.23~kJ.mol$^{-1}$~\cite{Cole83}.}
\label{param}
\end{table*}

\begin{table*}
\begin{center}
\begin{tabular}{|c|c|c|c|c|}
\hline
Electrolyte & Temperature (K) & N$_{\rm ions}$ & N$_{\rm ACN}$ & $L_z$ (nm) \\
\hline
[BMI][PF$_6$] & 400 & 320 & --- & 12.32 \\
\hline
[BMI][BF$_4$] & 400 & 320 & --- & 11.26 \\
\hline
ACN-[BMI][PF$_6$] & 298 & 96 & 896 & 12.27 \\
\hline
ACN-[BMI][BF$_4$] & 298 & 96 & 896 & 11.89 \\
\hline
\end{tabular}
\end{center}
\caption{Simulation temperature, number of ion pairs, number of ACN molecules and lengths of the simulation cell in the direction perpendicular to the graphite electrodes for the four electrolytes studied. The lengths in the $x$ and $y$ directions are the same for all the cells and are equal to 3.22~nm and 3.44~nm respectively.}
\label{cells}
\end{table*} 

Following our previous works~\cite{Merlet11,Merlet12b,Merlet12}, the electrodes are held at constant potential using a method developed by Reed {\it et al}~\cite{Reed07} from an original proposal by Siepmann and Sprik~\cite{Siepmann95}. A potential difference $\Delta\Psi^0$ is imposed between the positive ($\Psi^+$) and negative ($\Psi^-$) electrodes such that: $\Psi^+$~=~$-\Psi^-$~=~$\Delta\Psi^0/2$. Five potential differences between 0.0~V and 2.0~V are explored for the different electrolytes. The simulations are conducted starting with the 0.0~V potential difference and increasing it by steps of 0.5~V to facilitate the equilibration process. When the potential difference is increased, the system is allowed to equilibrate for at least 100~ps before collecting data for 1~ns. The constant potential approach is computationally expensive compared to the constant charge approach used in other works~\cite{Frolov12,Shim11,Shim12} but enables the analysis of the polarization of the electrodes by the electrolyte~\cite{Merlet12b,Merlet12}. Furthermore, this constant potential method may be applied to irregular porous electrodes~\cite{Merlet12}. 

\section{Results and discussion}
\label{Results}

\subsection{On the organization of the electrolyte at the interface}

\begin{figure*}
\includegraphics[scale=0.27]{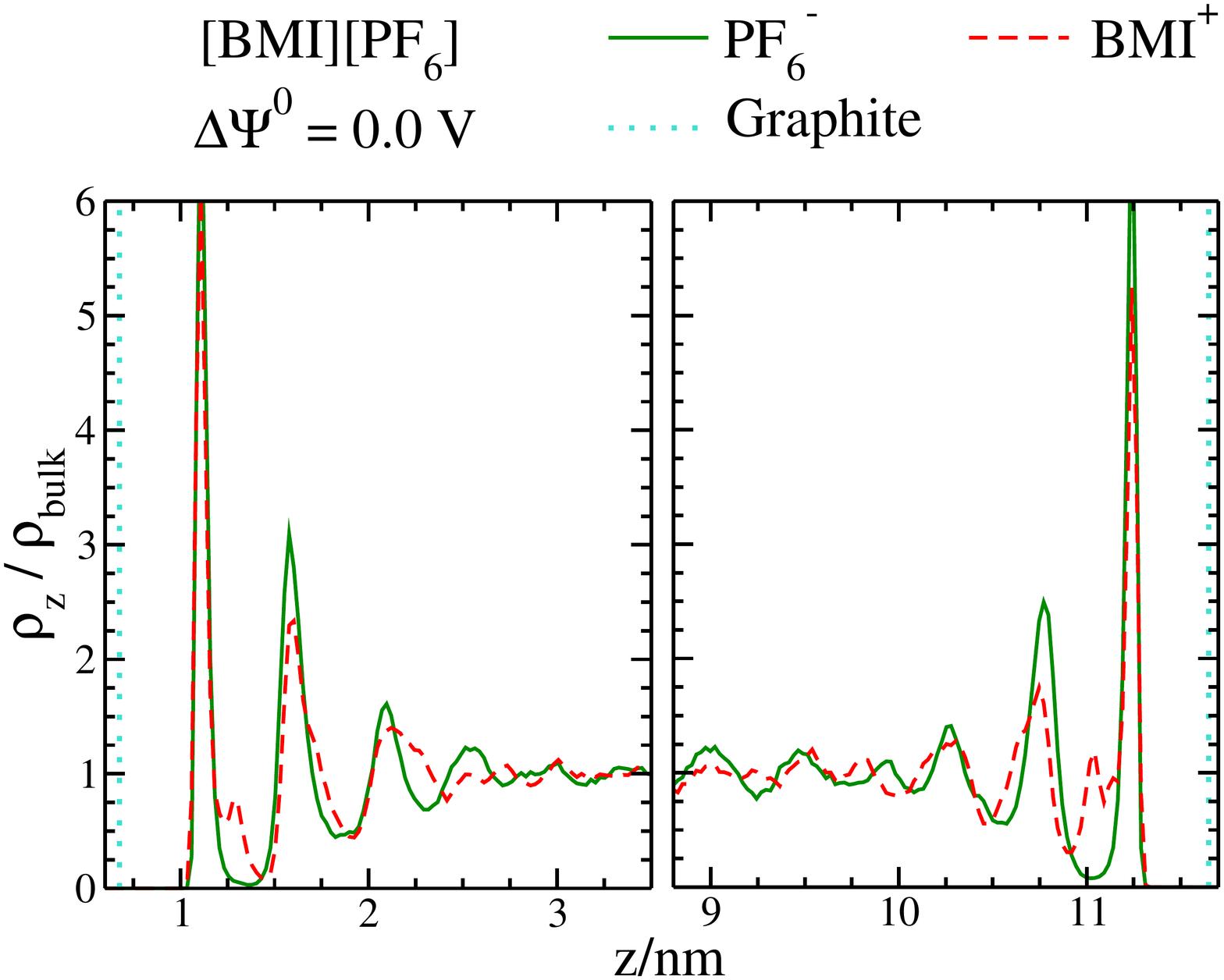}
\includegraphics[scale=0.27]{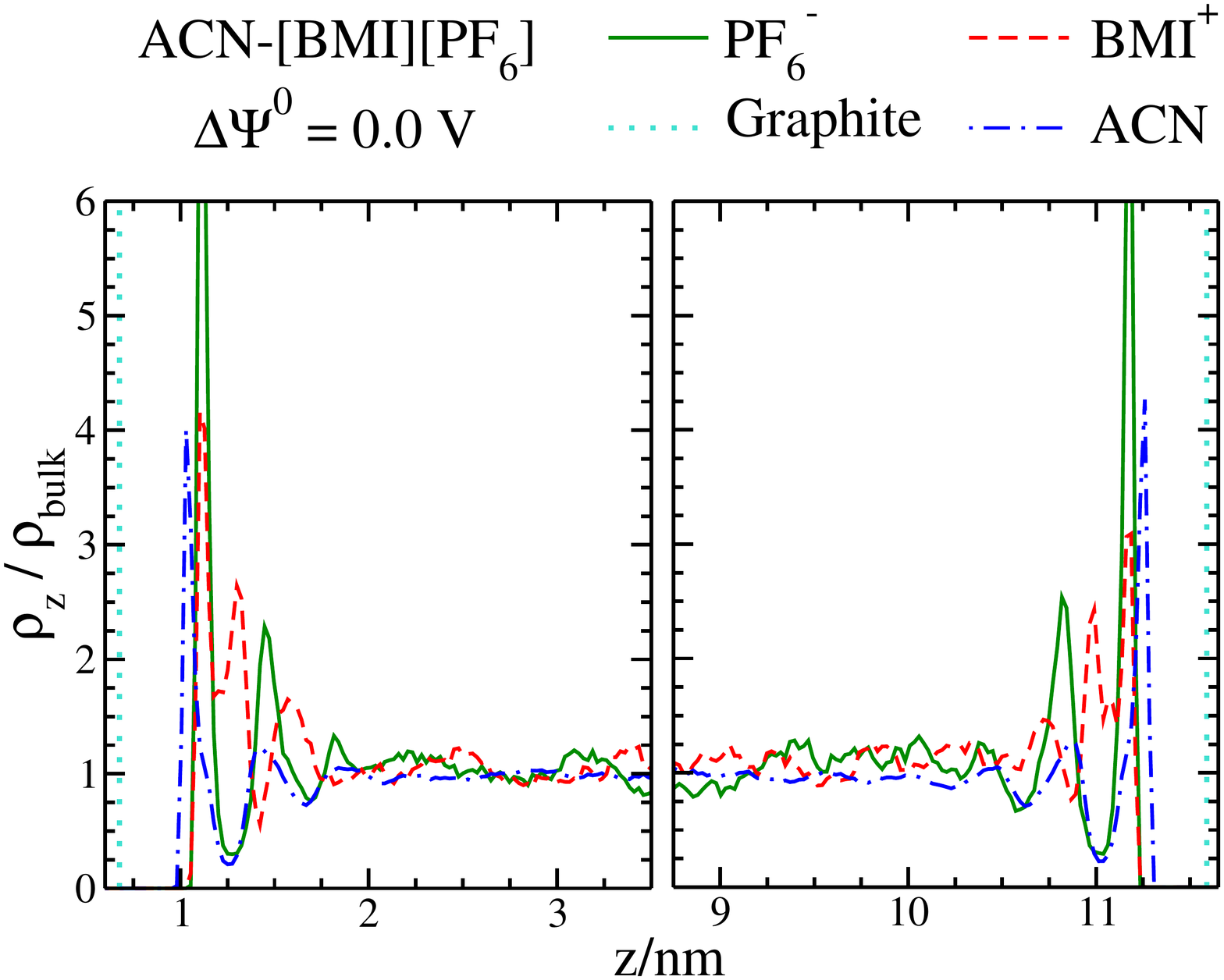}\\
\includegraphics[scale=0.27]{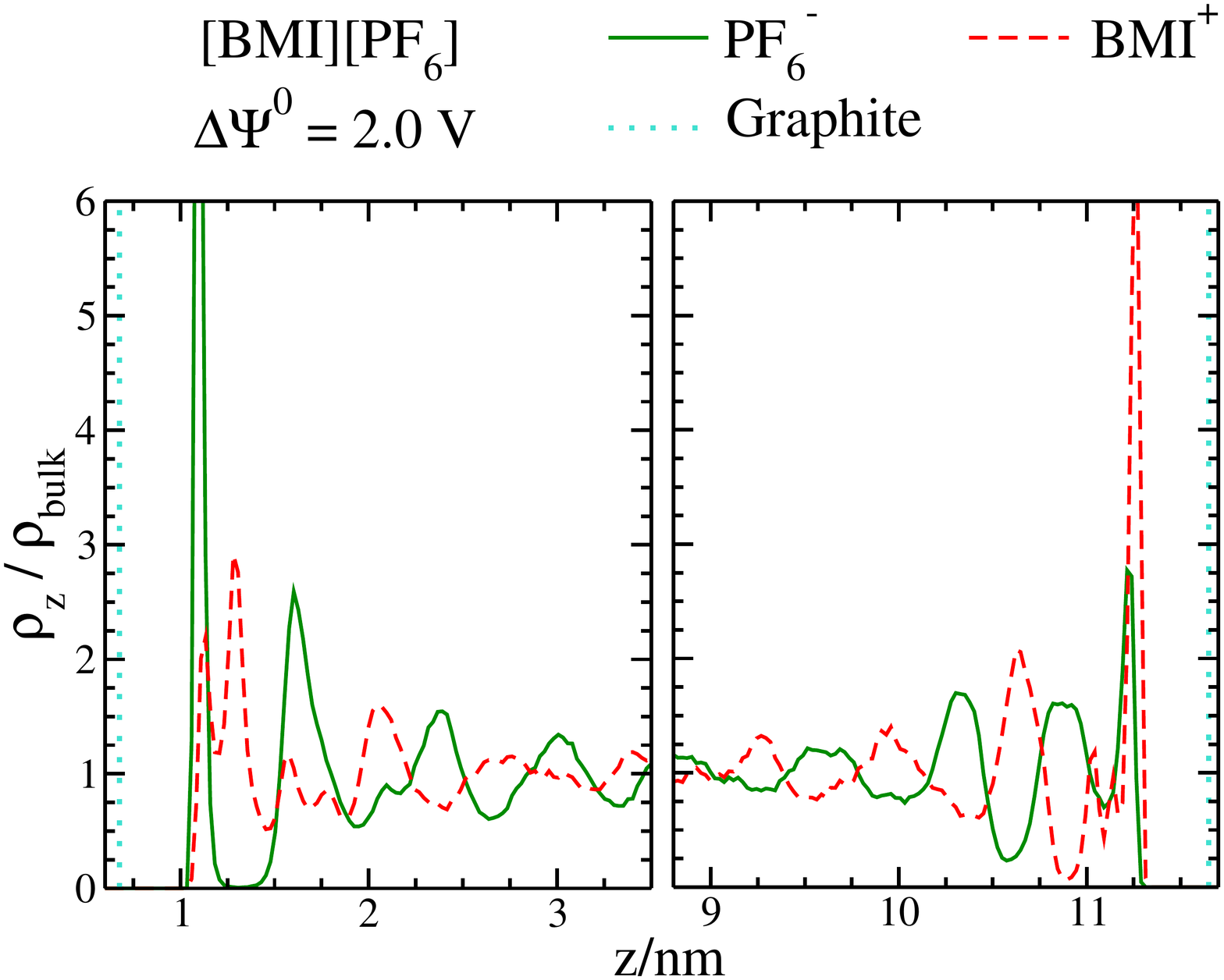}
\includegraphics[scale=0.27]{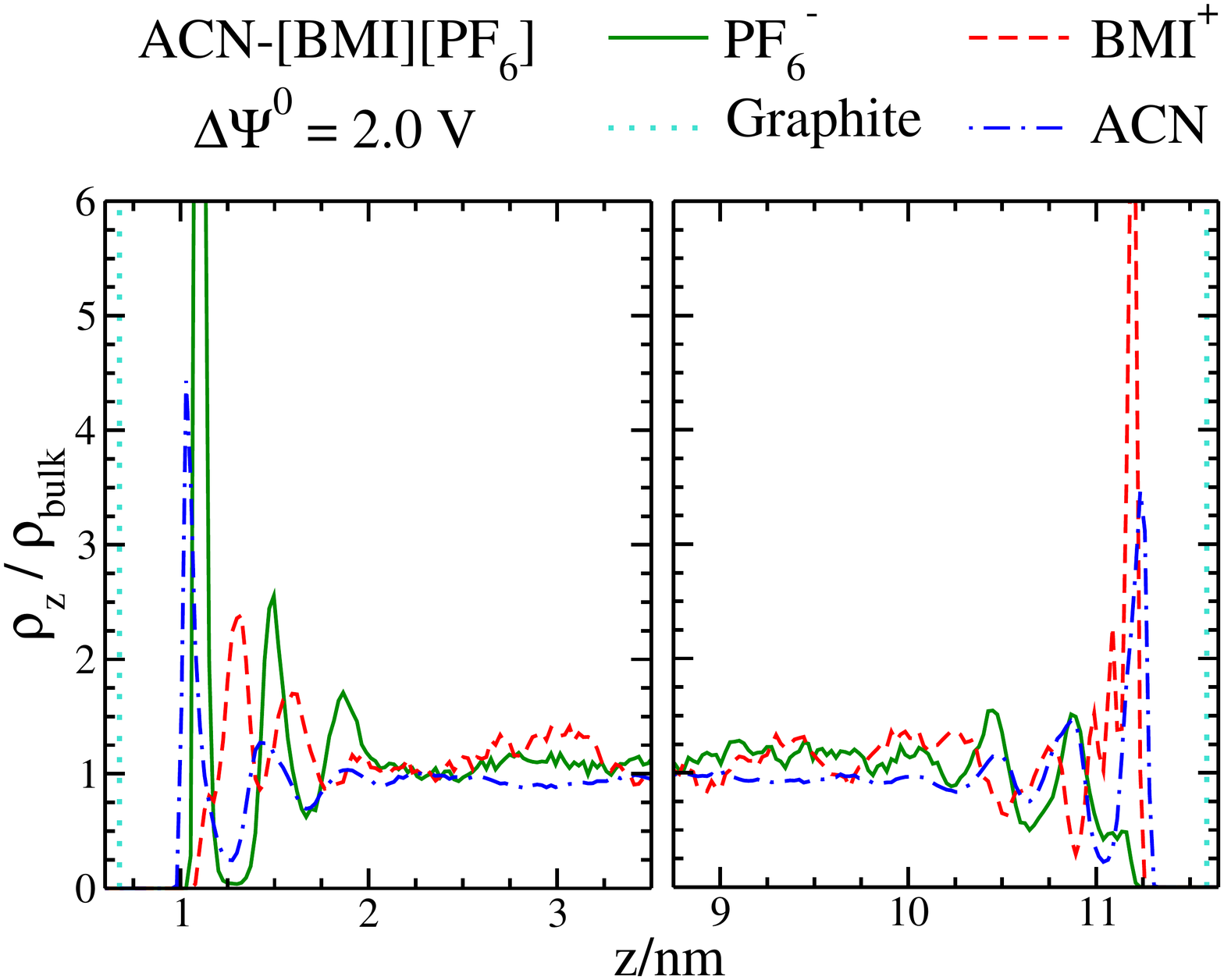}
\caption{Molecular densities of the center of mass of the different species, in the direction perpendicular to the electrodes, for pure [BMI][PF$_6$] and ACN-[BMI][PF$_6$]. Molecular densities are given for $\Delta\Psi^0$~=~0.0~V and $\Delta\Psi^0$~=~2.0~V. Blue dashed lines represent the positions of the graphite layers.}
\label{Density1}
\end{figure*}

\begin{figure*}
\includegraphics[scale=0.27]{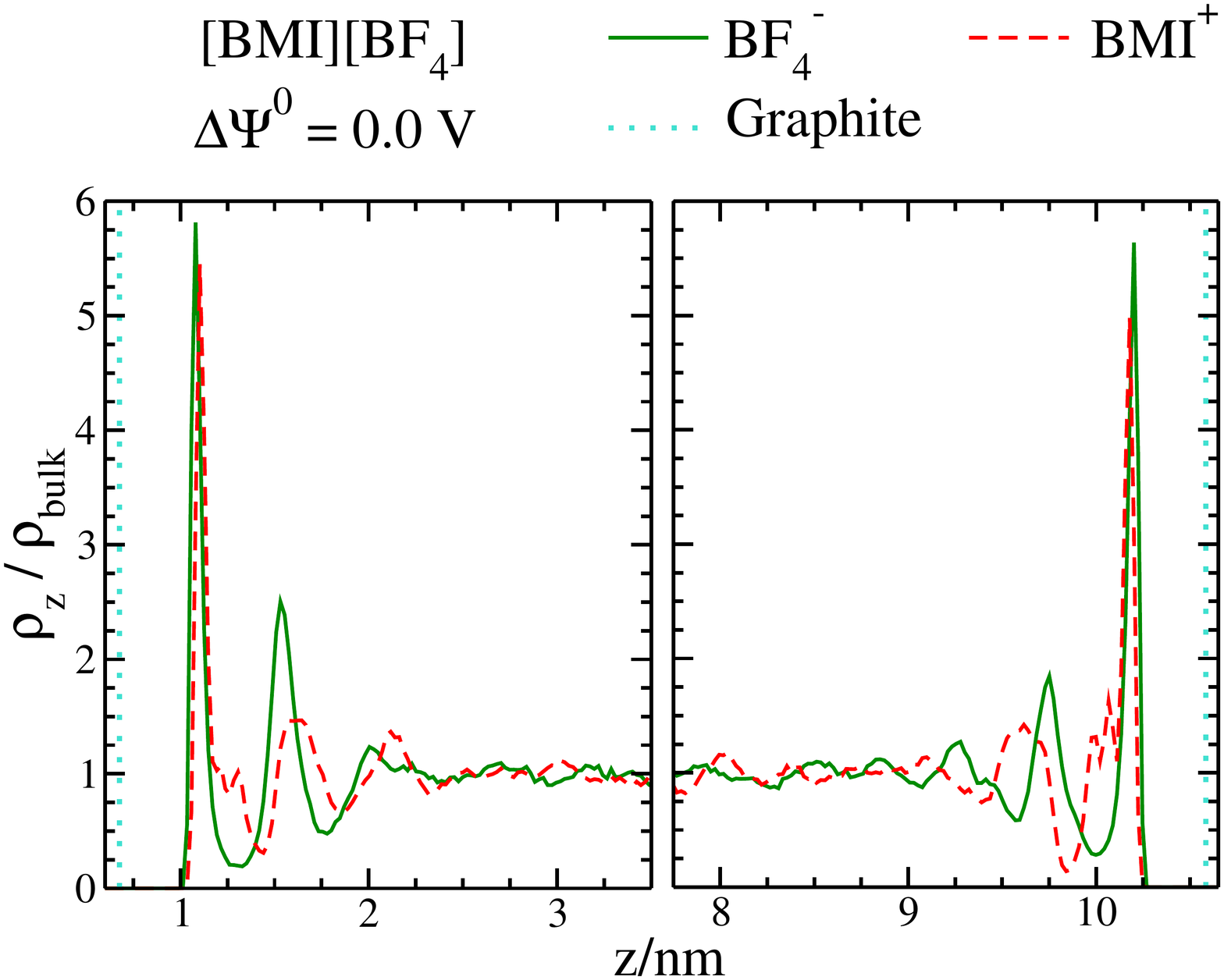}
\includegraphics[scale=0.27]{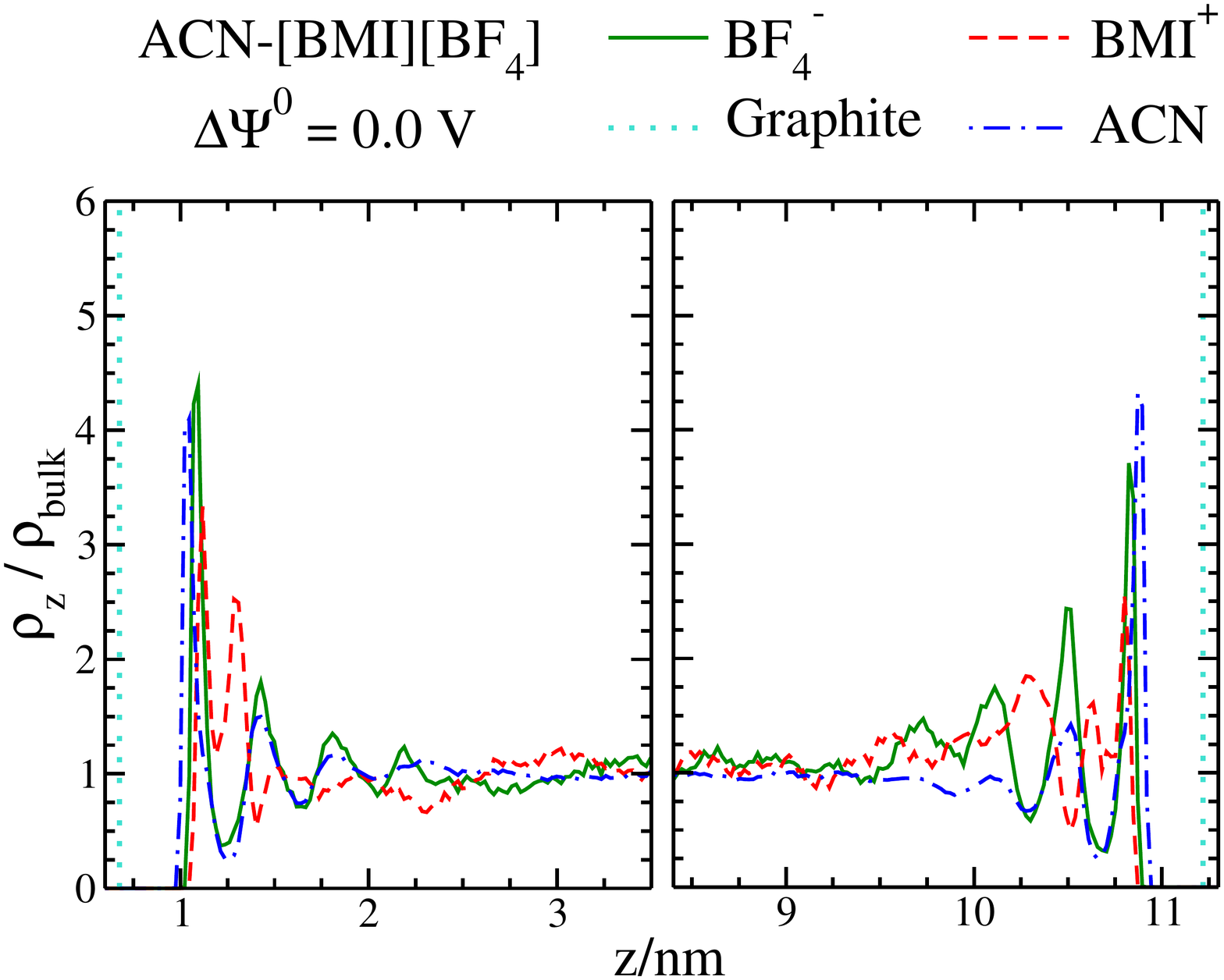}\\
\includegraphics[scale=0.27]{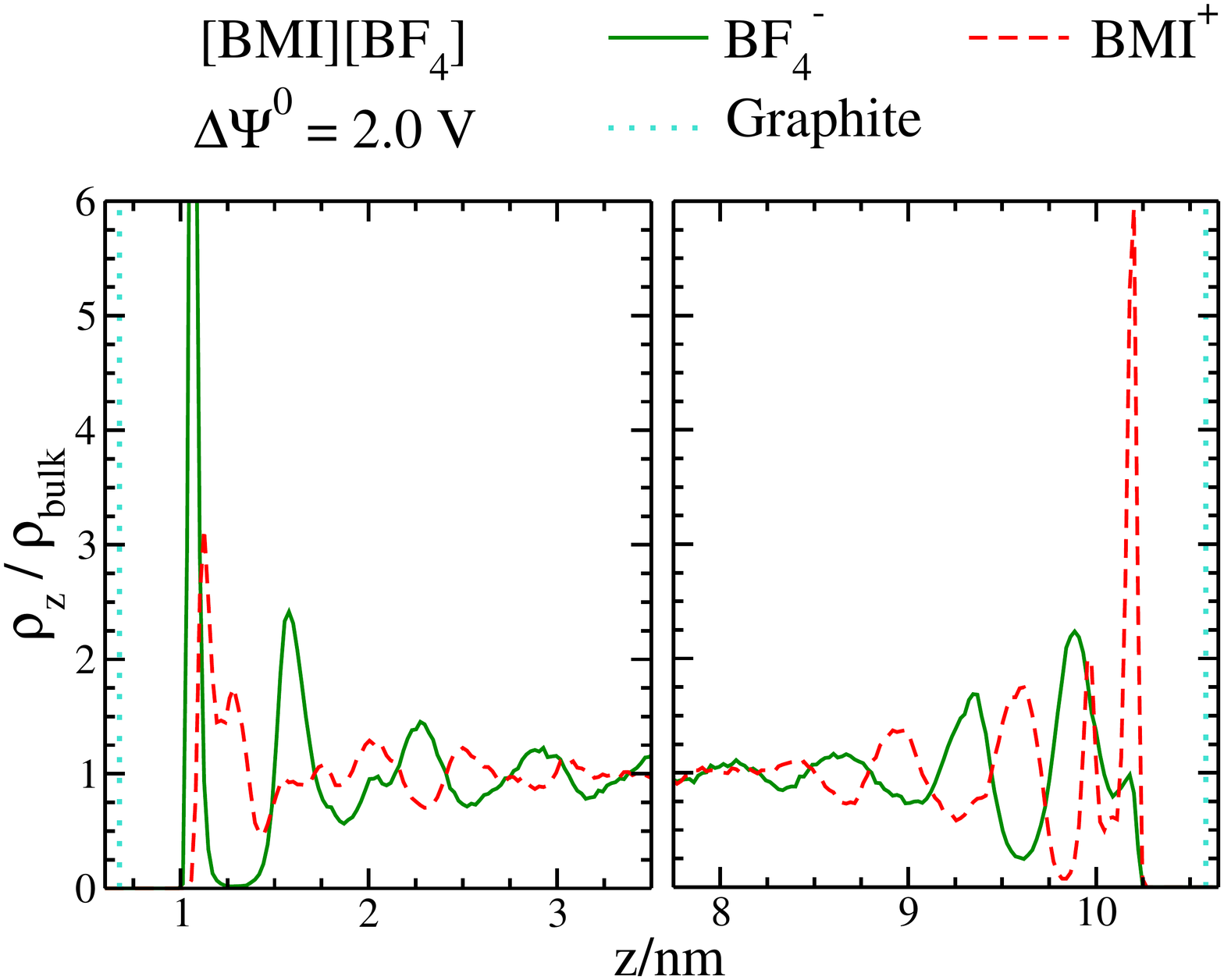}
\includegraphics[scale=0.27]{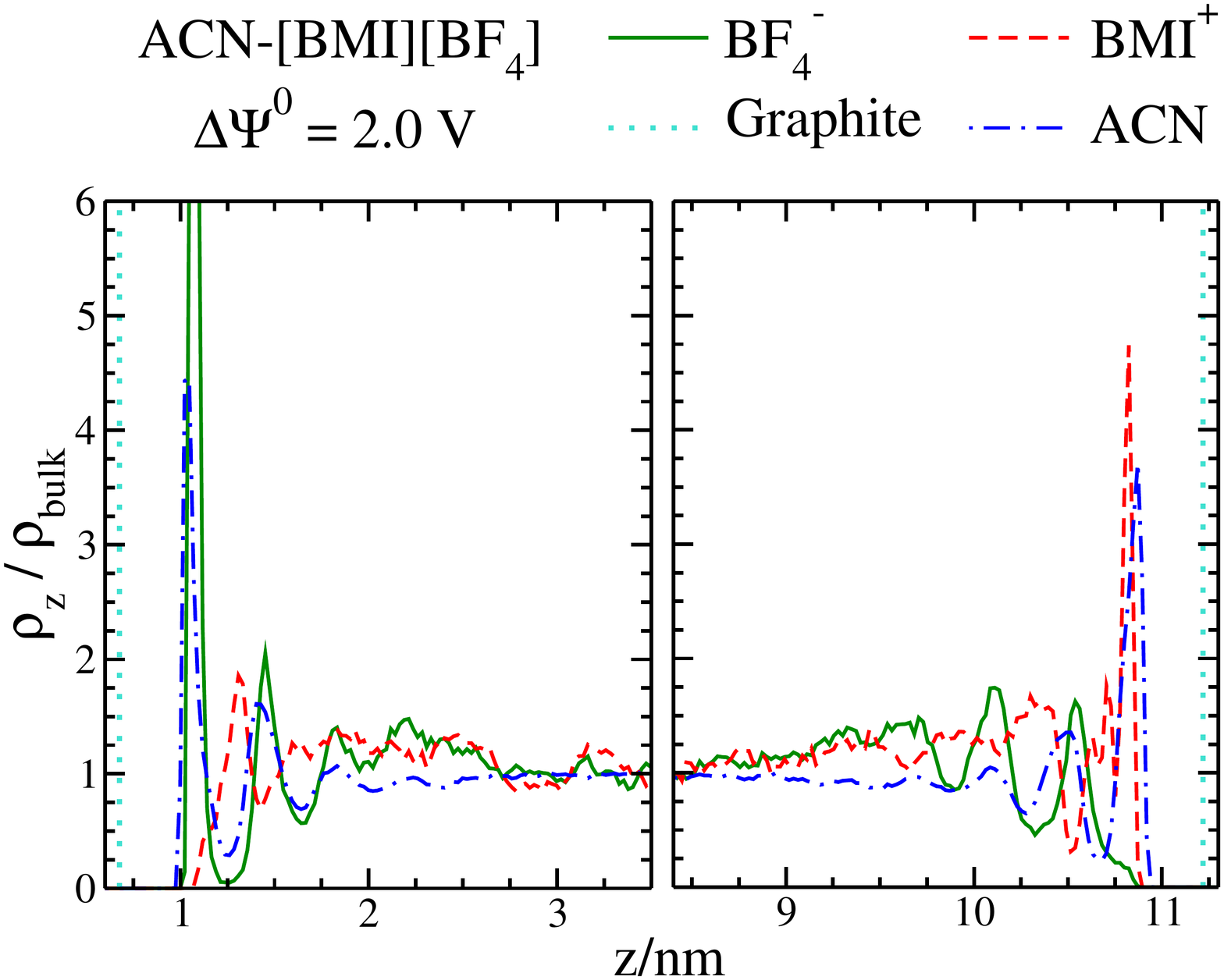}
\caption{Molecular densities of the center of mass of the different species, in the direction perpendicular to the electrodes, for pure [BMI][BF$_4$] and ACN-[BMI][BF$_4$]. Molecular densities are given for $\Delta\Psi^0$~=~0.0~V and $\Delta\Psi^0$~=~2.0~V. Blue dashed lines represent the positions of the graphite layers.}
\label{Density2}
\end{figure*}

The impact of the presence of solvent molecules on the structure of the electrolyte at the interface can be probed by computing molecular densities, $\rho_z$, in the direction perpendicular to the graphite electrodes. The molecular densities of the center of mass of the different species are represented in figure~\ref{Density1} (and~\ref{Density2}) for [BMI][PF$_6$] (and [BMI][BF$_4$]) based electrolytes. The calculated quantities for each species are normalized by the appropriate bulk densities to ease the comparison between the various electrolytes.

The first notable fact is the presence of molecular layering for all the species at the interface in both pure ILs and electrolyte solutions. This feature of the density profiles, which is well-known and has been observed for planar electrodes in both experiments~\cite{Atkin11,Hayes11} and simulations~\cite{Shim12,Merlet11,Merlet12b,Pounds09,Tazi10,Vatamanu10b}, cannot be recovered by mean-field theories which neglect ionic correlations, molecular sizes and fluctuations. We can underline here again that the classical Gouy-Chapman theory is not suited for studies of concentrated electrolytes.

The second important phenomenon, which is also present for all the electrolytes examined, is the reorganization of the layers upon charging. Alternating layers of ions of opposite charge are associated with the so-called overscreening effect~\cite{Feng11b,Reed07,Lanning04,Bazant11,Heyes81,Esnouf88}: The charge in the first adsorbed layer overcompensates that on the electrodes, and, due to correlation between ions, the residual charge is successively overcompensated by the charge in the second adsorbed layer and so on, until the bulk density is reached. At this point, we can underline the first notable consequence of solvation which is to reduce the region where density oscillations are visible from a thickness of around 2~nm for pure ILs to approximately 1~nm for electrolyte solutions. More precisely, in the electrolyte {\emph solutions}, the first adsorbed layer of counterions overcompensates the charge on the electrode but the overscreening effect does not go beyond two molecular layers. This reduction of the overscreening can be highlighted by plotting the integral of the charge density of the liquid normalized by the the electrode surface charge as described by Feng \emph{et al.}~\cite{Feng11b}. Our results (not shown here for brevity) are in qualitative agreement with molecular dynamics simulation of the [BMI][BF$_4$] and ACN-[BMI-BF$_4$] mixtures, near graphite electrodes, for which it is shown that both the intensity and extension of the overscreening effect are decreased in electrolyte solutions in comparison with neat ionic liquids~\cite{Feng11b}. 

\begin{figure*}
\begin{center}
\includegraphics[scale=0.60]{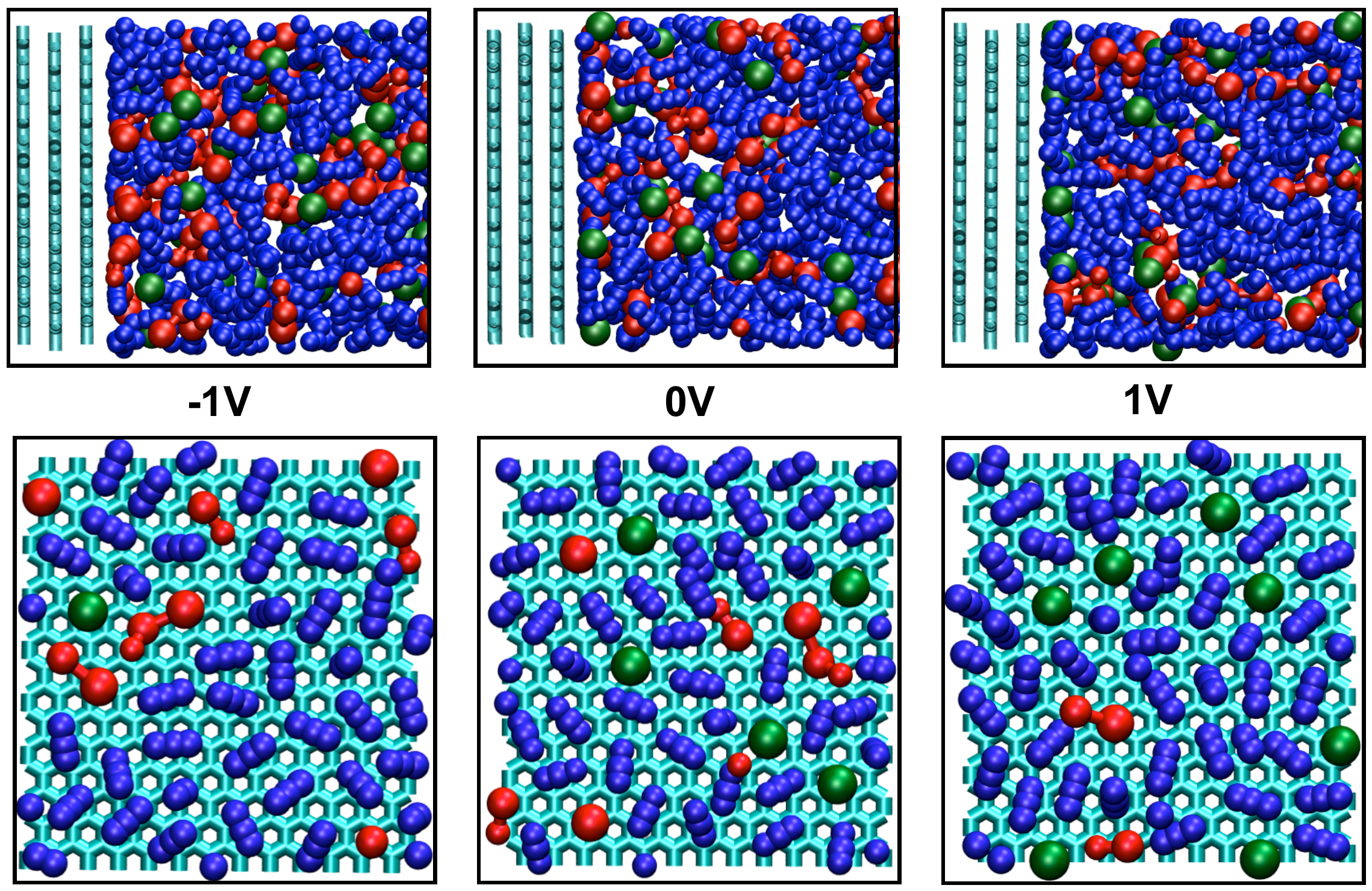}
\end{center}
\caption{Snapshots of the interface: side view (upper pannel) and top view (lower pannel). Snapshots are given for different potentials: 1.5~M solution of BMI$^+$ cations (red) and BF$_4^-$ anions (green) in ACN (dark blue) near graphite walls (light blue). For all potentials, inner sphere adsorbed ions are visible.}
\label{SnapNV}
\end{figure*}

Going further into the analysis of the molecular densities, it appears that the density peaks of the first adsorbed ionic layers are enhanced in [BMI][PF$_6$] based electrolytes compared to [BMI][BF$_4$] based electrolytes. The positions of the first peaks in the ionic density profiles are not shifted away from the electrode upon addition of a solvent, leading to the conclusion that tightly adsorbed ions exist at the interface. This situation can be visualized in snapshots of the simulations (see figure~\ref{SnapNV}). In a given electrolyte, the ionic species also show different affinities for the graphite surface, the resulting dissymmetry in the density profiles being larger in the ACN containing mixtures. A common feature is the important variation of the heights and positions of the ionic density peaks when the potential difference is changed. 

On the contrary, the molecular density of ACN depends neither on the nature of the anion nor on the potential difference applied. As a consequence, while in pure ILs, the potential difference increase induces mainly a polarization of the layers near the electrodes, in electrolyte solutions, exchanges of ions between different layers occur. One should keep in mind that the ions, in the 1.5~M solutions, represent only 10~\% of the molecules. Thus, even if the ionic densities are modified upon charging, the major component at the interface and in the bulk is the solvent which can probably accommodate the charging of the electrode by rearranging only a small number of molecules. On the positive electrode side, this induces a structure where the counter-ions and ACN molecules are located in the same plane (see figure~\ref{SnapNV}). The cations are simply reoriented and the solvation shell of the ions is slightly distorted in the vicinity of the graphite surface. On the negative electrode side, the positions of the anions are shifted away from the graphite surface but still lie in the same plane as the ACN molecules. 

Another notable consequence of the presence of solvent is the stronger expulsion of co-ions from the first adsorbed layer for the 2~V potential difference. Indeed, in pure ILs for $\Delta\Psi$~=~2.0~V, the heights of the co-ion peaks of the first layers are reduced by a factor between 2 and 6 but peaks are still apparent. For the same potential difference in electrolyte solutions, co-ions are almost absent in the first adsorbed layer. This suggests that the removal of co-ions is facilitated in the electrolyte solution.

\begin{figure*}
\includegraphics[scale=0.27]{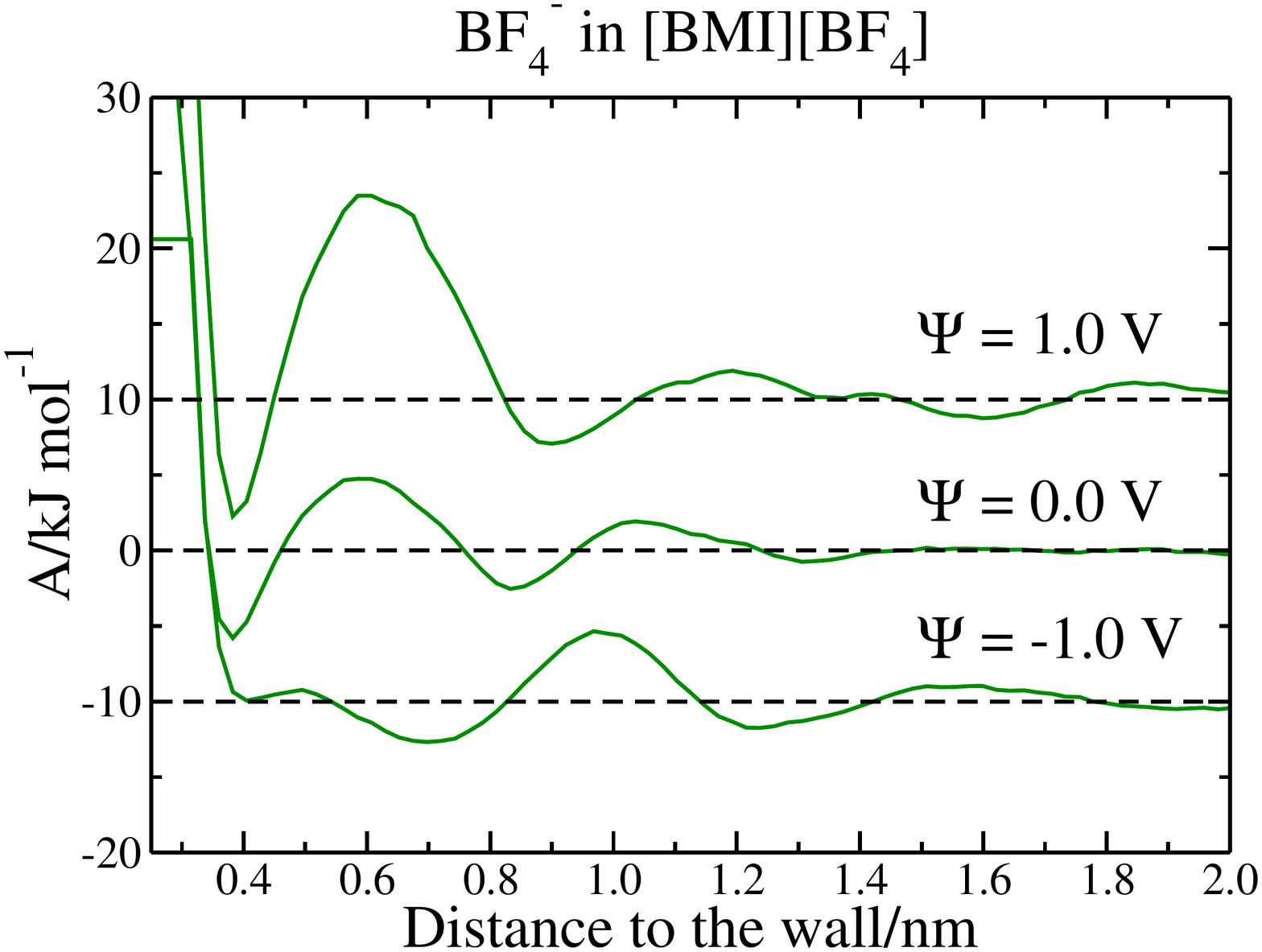}
\includegraphics[scale=0.27]{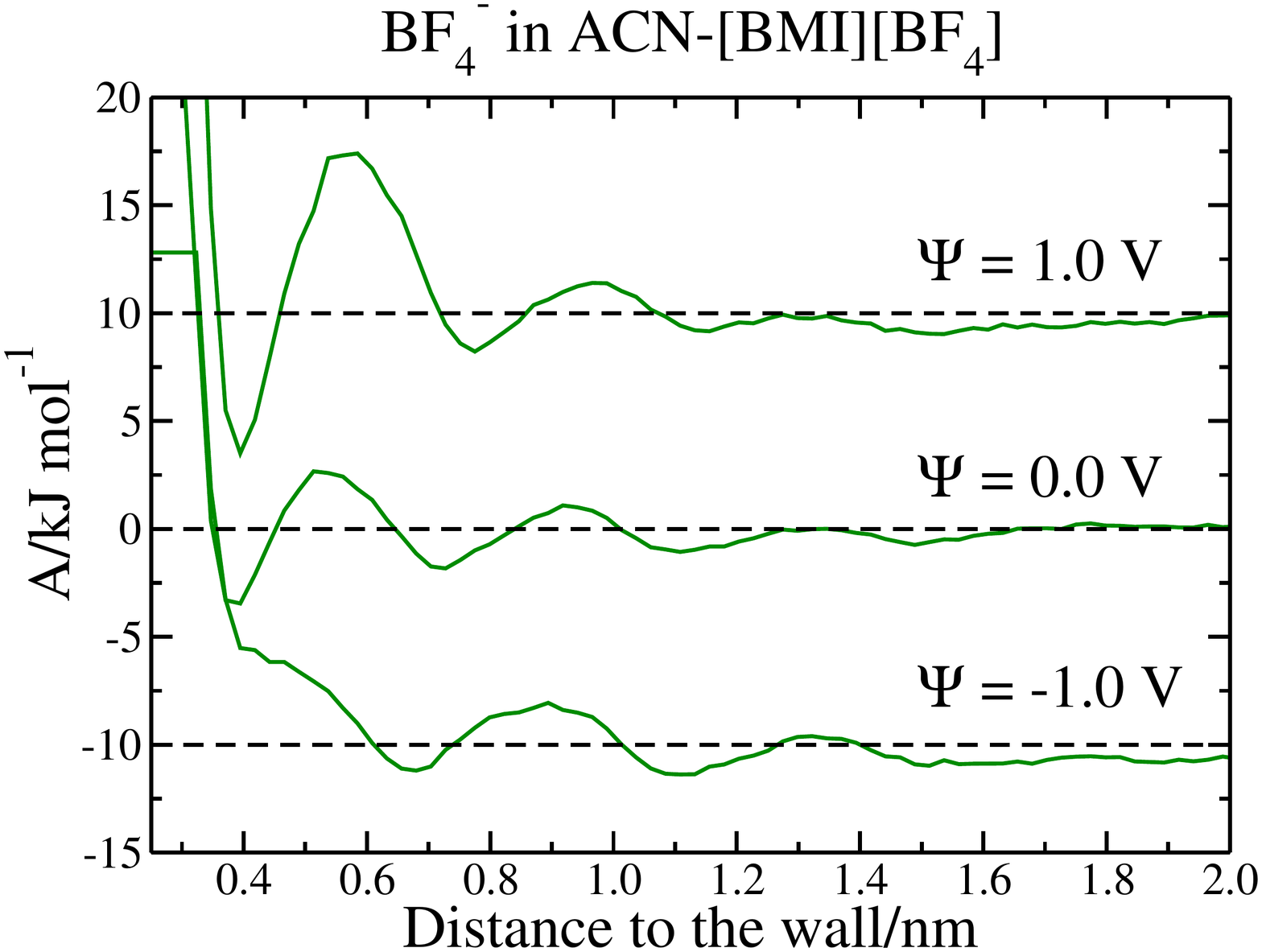}\\
\includegraphics[scale=0.27]{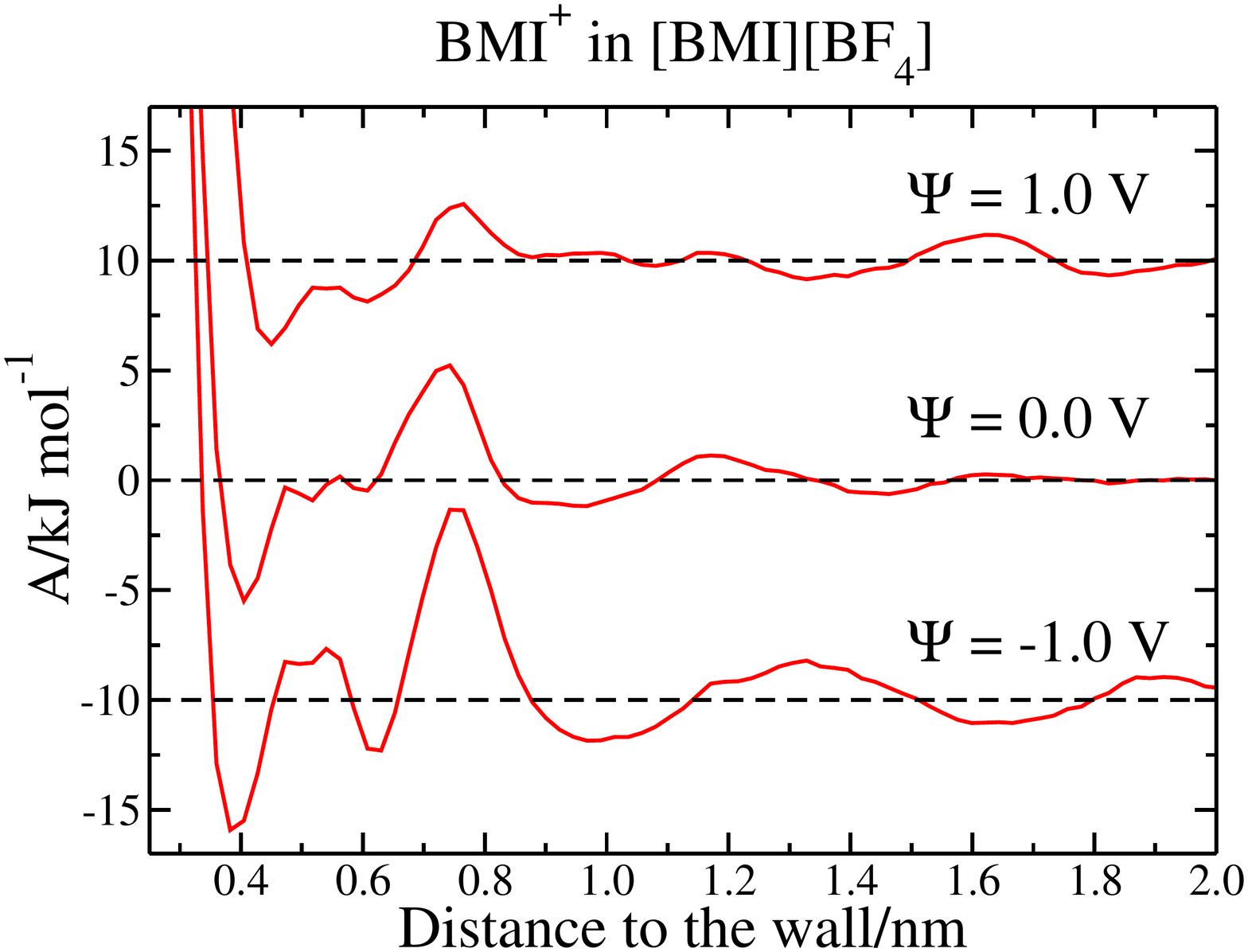}
\includegraphics[scale=0.27]{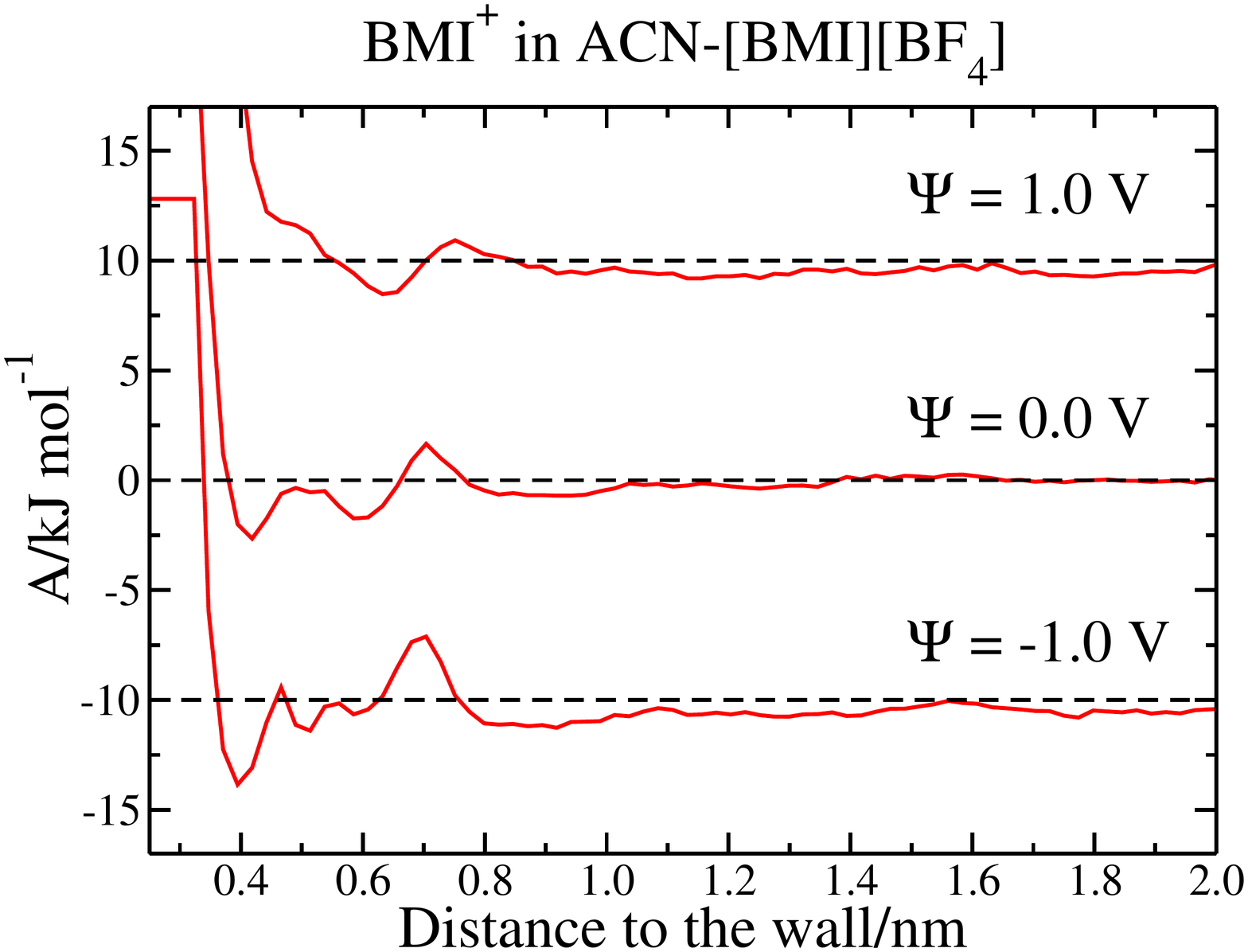}
\caption{Free energy profiles for anions and cations in the [BMI][BF$_4$] and ACN-[BMI][BF$_4$] electrolytes. The curves for the negative and positive electrodes are shifted by -10~kJ.mol$^{-1}$ and 10~kJ.mol$^{-1}$ respectively for visualization purposes.}
\label{free_eng}
\end{figure*}

Another way of looking at these results is to plot the free energy profiles of the various species:
\begin{equation}
A(z) - A_{bulk} = -kT\ln(\frac{\rho_z}{\rho_{bulk}}),
\end{equation}
where $k$ is the Boltzmann constant, $T$ is the temperature of the system and $\rho_z$/$\rho_{bulk}$ is the molecular density at a given position normalized by the bulk density. The free energy profiles for the anions and cations in the [BMI][BF$_4$] and ACN-[BMI][BF$_4$] electrolytes are shown in figure~\ref{free_eng}. For the zero potential difference, all free energy profiles are characterized by a well near the wall at a distance of approximately 0.4~nm. For a 2~V potential difference, the energy barrier to overcome for an ion to go from the bulk to the first adsorbed layer increases for a favorably charged surface and decreases for an unfavorably charged surface. The same observation was made from simulations of 1,3-dimethylimidazolium chloride near graphite walls using positive, negative and neutral probes~\cite{Lynden-Bell12}. In the case of unfavorably charged surfaces, this well is still visible in the solvent-free electrolyte, even if very small in the case of BF$_4^-$, but missing in ACN-[BMI][BF$_4$]. The curves for the [BMI][PF$_6$] based electrolytes (not shown) lead to the same conclusions. The fact that co-ions are expelled from the first adsorbed layer more easily in electrolyte solutions, in comparison to pure ILs, is linked with the decrease of ion-ion correlations upon addition of a solvent and is consistent with a charging mechanism where ions can be exchanged between different layers at the interface.

The next step in the structural analysis is the calculation of coordination numbers in the bulk and at the interface for various potential differences. The coordination number for each species was estimated as the average number of molecules at a distance smaller than the first minimum in the appropriate bulk radial distribution function. The coordination numbers at the interface were computed as the average coordination number for molecules located in the first adsorbed layer. All the computed values are summarized in table~\ref{coord}.

\begin{table*}
\begin{center}
\begin{tabular}{|c|c|c|c|c|c|}
\hline
Electrode Potential & Electrolyte & $N_{\rm C}({\rm A})$ & $N_{\rm A}({\rm C})$ & $N_{\rm ACN}({\rm A})$ & $N_{\rm ACN}({\rm C})$\\
\hline
\multirow{4}{*}{Bulk} & [BMI][PF$_6$] & 6.0 & 6.0 & -- & -- \\
 & ACN-[BMI][PF$_6$] & 1.8 & 1.8 & 9.3 & 6.7 \\
 & [BMI][BF$_4$] & 6.0 & 6.0 & -- & -- \\
 & ACN-[BMI][BF$_4$] & 1.9 & 1.9 & 8.8 & 6.7 \\
\hline
\multirow{4}{*}{0.0~V} & [BMI][PF$_6$] & 5.0 & 5.0 & -- & -- \\
 & ACN-[BMI][PF$_6$] & 1.6 & 1.6 & 7.0 & 5.0 \\
 & [BMI][BF$_4$] & 4.8 & 4.8 & -- & -- \\
 & ACN-[BMI][BF$_4$] & 1.4 & 1.4 & 7.2 & 5.2 \\
\hline
\multirow{4}{*}{-1.0~V} & [BMI][PF$_6$] & 5.1 & 4.0 & -- & -- \\
 & ACN-[BMI][PF$_6$] & 2.1 & 0.9 & 6.7 & 5.5 \\
 & [BMI][BF$_4$] & 4.9 & 3.9 & -- & -- \\
 & ACN-[BMI][BF$_4$] & 2.0 & 0.8 & 6.2 & 5.1 \\
\hline
\multirow{4}{*}{1.0~V} & [BMI][PF$_6$] & 4.6 & 5.4 & -- & -- \\
 & ACN-[BMI][PF$_6$] & 1.2 & 2.2 & 7.7 & 5.2 \\
 & [BMI][BF$_4$] & 4.6 & 5.5 & -- & -- \\
 & ACN-[BMI][BF$_4$] & 0.9 & 2.1 & 7.9 & 5.3 \\
\hline
\end{tabular}
\end{center}
\caption{Coordination numbers in the bulk and at the interface for anions and cations. The coordination number at the interface is the average coordination number for molecules located in the first adsorbed layer. $N_{\rm i}({\rm j})$ is the number of molecules of species i surrounding molecules of species j.}
\label{coord}
\end{table*}
Firstly, we focus on the counter-ion coordination numbers around a given ion. At 0V, we observe a systematic decrease of this coordination number at the interface compared to the bulk, due to the proximity of the carbon atoms of the electrode surface. This reduction is more pronounced in the pure ILs (-1.0 to -1.2 units) compared to the ACN-based electrolytes (-0.2 to -0.5 units), with an increased effect when the anion is BF$_4^-$. As soon as a positive (respectively negative) potential is applied, the anions (cations) coordination sphere tends to diminish further due to the presence of compensating charge at the surface of the electrode. Equally, the cation (anion) interaction with the carbon now needs to be screened, provoking an increase of the coordination number, which can even go above the bulk one in the case of electrolyte solutions. 

Secondly, we consider the coordination numbers of ACN around ions for electrolyte solutions which can be referred to as solvation numbers. An interesting feature is the larger decrease of coordinating ACN compared to coordinating ions at the interface. This shows that it is somewhat easier to remove ACN from the coordination shell compared to ions. We note that, upon polarization, very slight changes are observed in the ACN coordination numbers around ions. 
 
We can also see that the PF$_6^-$ ions are more readily desolvated than the BF$_4^-$ ions. This statement is consistent with the Born model which describes the Gibbs energy of the ion-solvent interaction as:
\begin{equation}
\Delta G_{IS} = -\frac{z^2e^2N_a}{8\pi\varepsilon_0r_{ion}} \times \left (1-\frac{1}{\varepsilon_r} \right ),
\end{equation}
where $z$ is the valence of the ion, $e$ is the elementary charge, $N_a$ is Avogadro constant, $\varepsilon_0$ and $\varepsilon_r$ are respectively the permittivity of vacuum and solvent. When the ion size is smaller, the interaction energy increases and the desolvation is more difficult.

\subsection{On the polarization of the electrodes}

\begin{figure*}
\includegraphics[scale=0.27]{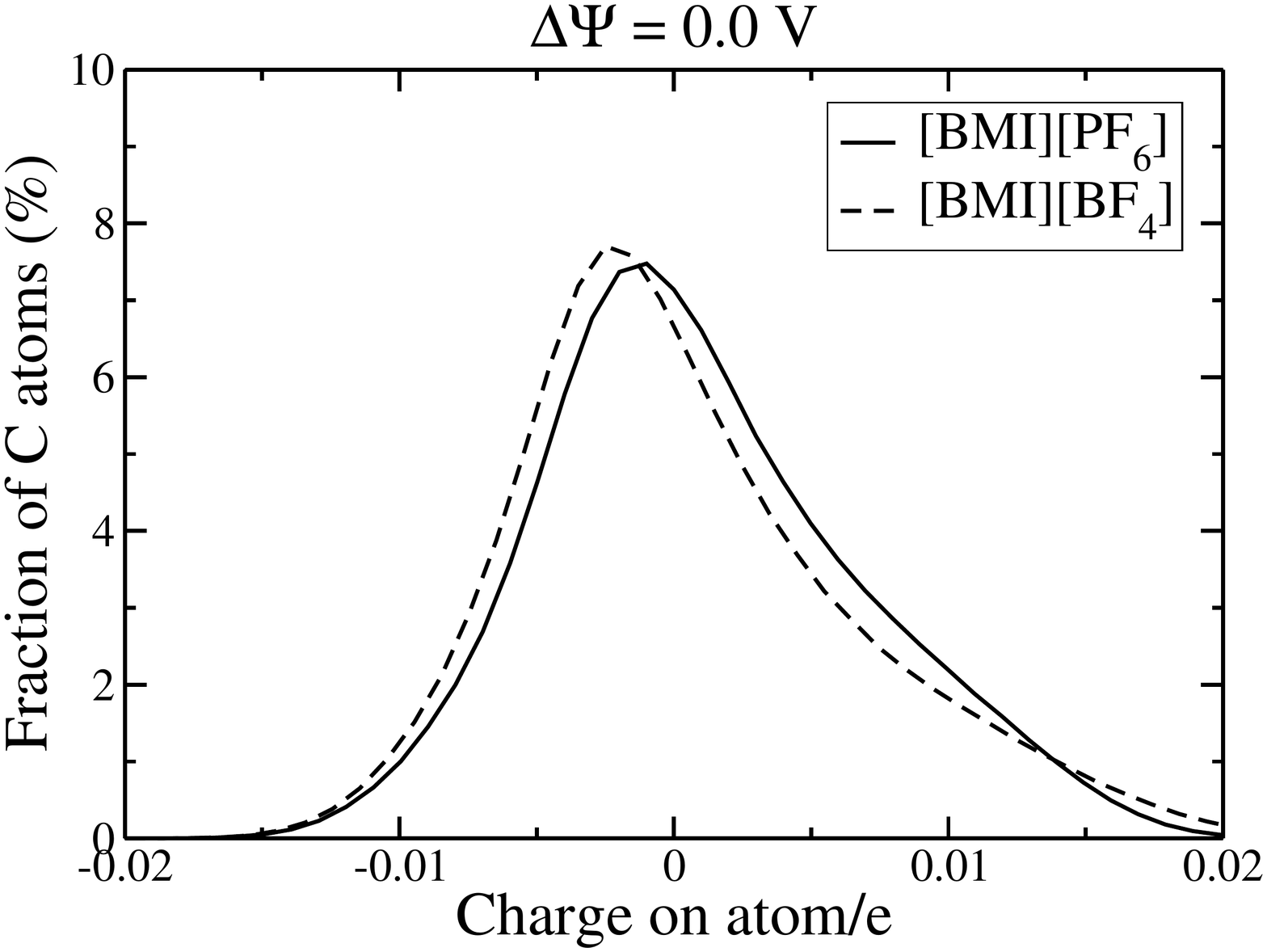}
\includegraphics[scale=0.27]{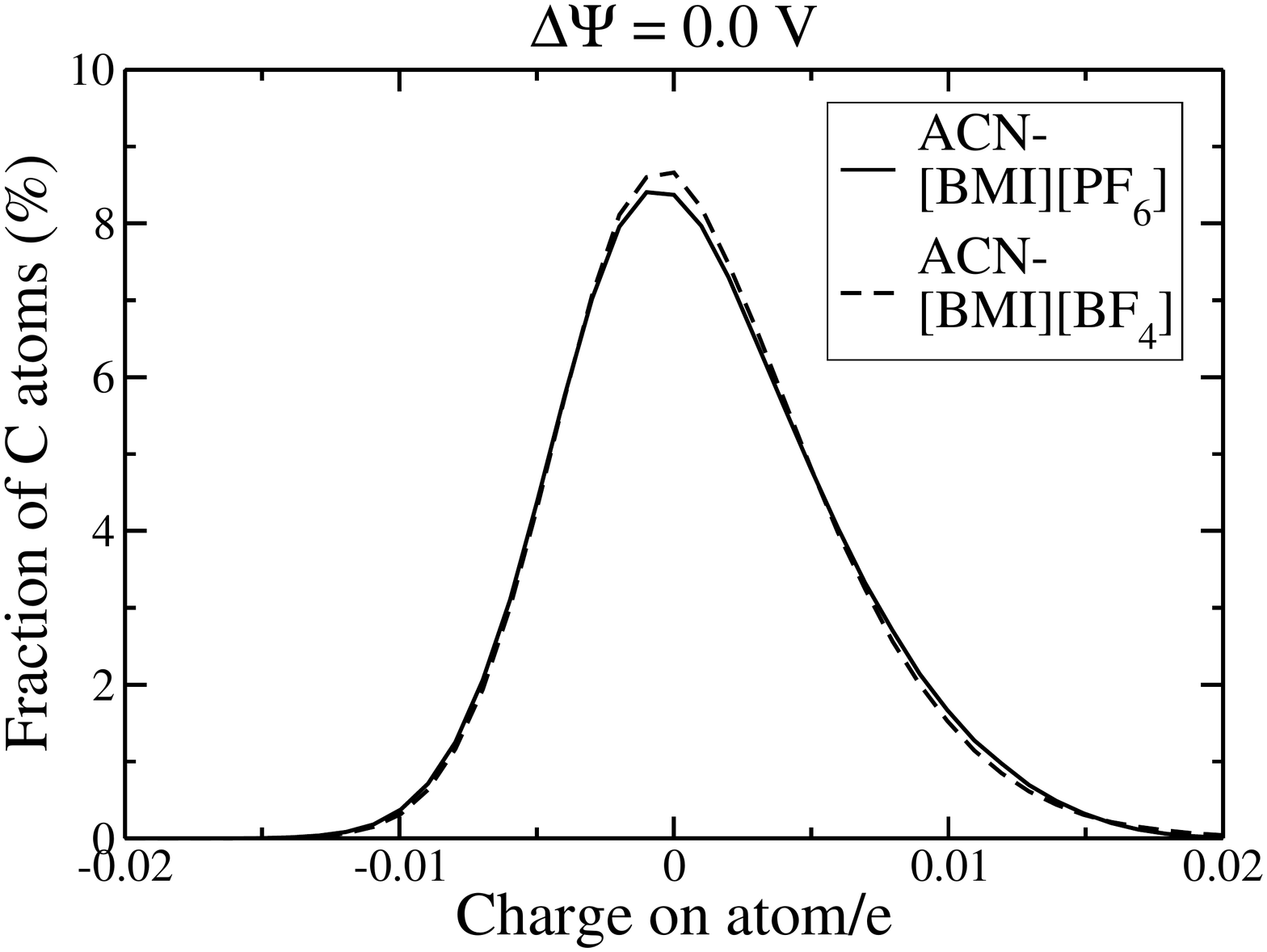}\\
\includegraphics[scale=0.27]{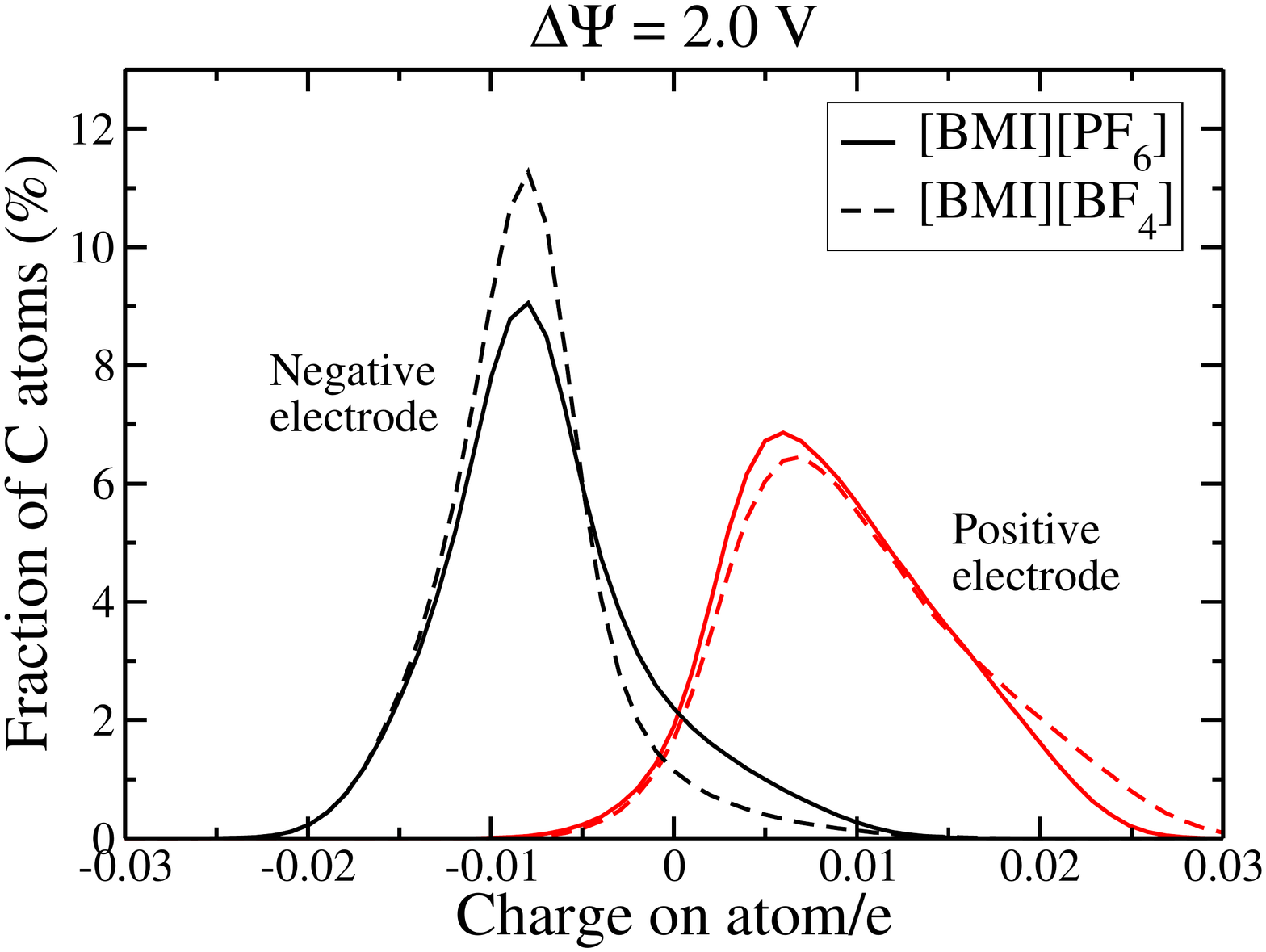}
\includegraphics[scale=0.27]{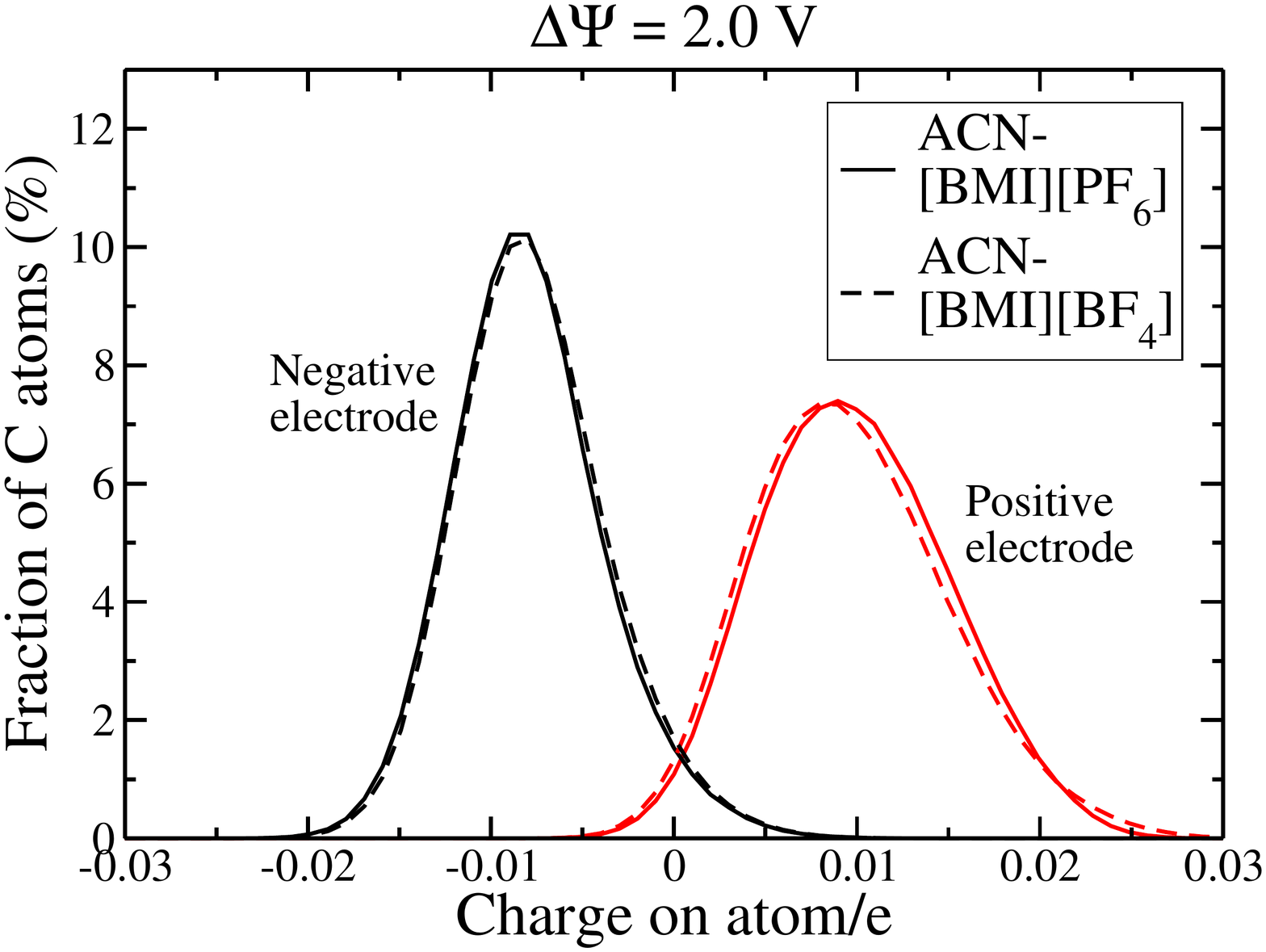}
\caption{Distributions of the carbon charges on the electrodes for the different liquids. Results for $\Delta\Psi^0$~=~0.0~V and $\Delta\Psi^0$~=~2.0~V are shown. When mixing the salts with acetonitrile as a solvent, the charge distribution curves become very similar.}
\label{histo}
\end{figure*}

Our constant potential approach for modelling electrochemical systems allows us to gain insight into the influence of the presence of solvent molecules on the polarization of the electrodes. The charges on the electrode atoms fluctuate during the simulations and it is possible to plot charge distribution functions, i.e. the fraction of carbon atoms that have a given charge. Charge distributions for the studied electrolytes and various potential differences are shown in figure~\ref{histo} (the analysis is focused on the first graphene layer near the electrolyte). The mean charge and the charge corresponding to the maximum occurrence are detailed in table~\ref{mean_max}. 

\begin{table*}
\begin{center}
\begin{tabular}{|c|c|c|}
\hline
Electrolyte & Mean charge (e) & Most frequent charge (e) \\
\hline
[BMI][PF$_6$] & $\pm$~8.02~$10^{-3}$ & +~5.97~$10^{-3}$ / -~7.96~$10^{-3}$ \\
\hline
[BMI][BF$_4$] & $\pm$~9.10~$10^{-3}$ & +~6.96~$10^{-3}$ / -~7.96~$10^{-3}$ \\ 
\hline
ACN-[BMI][PF$_6$] & $\pm$~8.91~$10^{-3}$ & +~8.96~$10^{-3}$ / -~8.40~$10^{-3}$ \\ 
\hline
ACN-[BMI][BF$_4$] & $\pm$~8.65~$10^{-3}$ & +~8.47~$10^{-3}$ / -~8.45~$10^{-3}$ \\ 
\hline
\end{tabular}
\end{center}
\caption{Mean charge and most frequent charge in the first graphite layer of the electrode, i.e. the layer closest to the electrolyte, for the potential difference $\Delta\Psi^0$~=~2~V.}
\label{mean_max}
\end{table*}
The presence of solvent seems to have two effects on the polarization of the electrodes. The first one is a decrease in the skewness of the charge distributions shapes. When the distribution is less skewed, the charge corresponding to the most frequent charge is closer to the average charge. Secondly, going from pure ILs to electrolyte solutions, the charge distribution functions become very similar when passing from [BMI][PF$_6$] to [BMI][BF$_4$] and can be superimposed. The difference between mean charges is reduced from 13~\% between pure ILs to 3~\% between electrolyte solutions. 

Independently of the nature of the electrolyte, there is a broadening of the charge distribution functions going from negative potentials to zero potentials to positive potentials. This may be attributed to the asymmetry between anions and cations in the ILs with the smaller anions inducing larger local positive charges on the graphite. When solvent is present, the orientation of the dipolar ACN molecules changes slightly with the sign of the electrode potential, which can induce different polarization at the surface. Our constant potential calculations thus reveal the importance of several factors on the polarization of the electrodes: i) The potential difference applied generates a different environment near the electrode and a broadening/contraction of the distribution curve, ii) In pure ILs, the nature of the anion has an influence on the shape of the curves, iii) The solvation of salts induces a shift and symmetrization of the charge distribution curves, and reduces the impact of the size of the anion.

\subsection{On the capacitive behavior of the system}

\begin{figure*}
\includegraphics[scale=0.27]{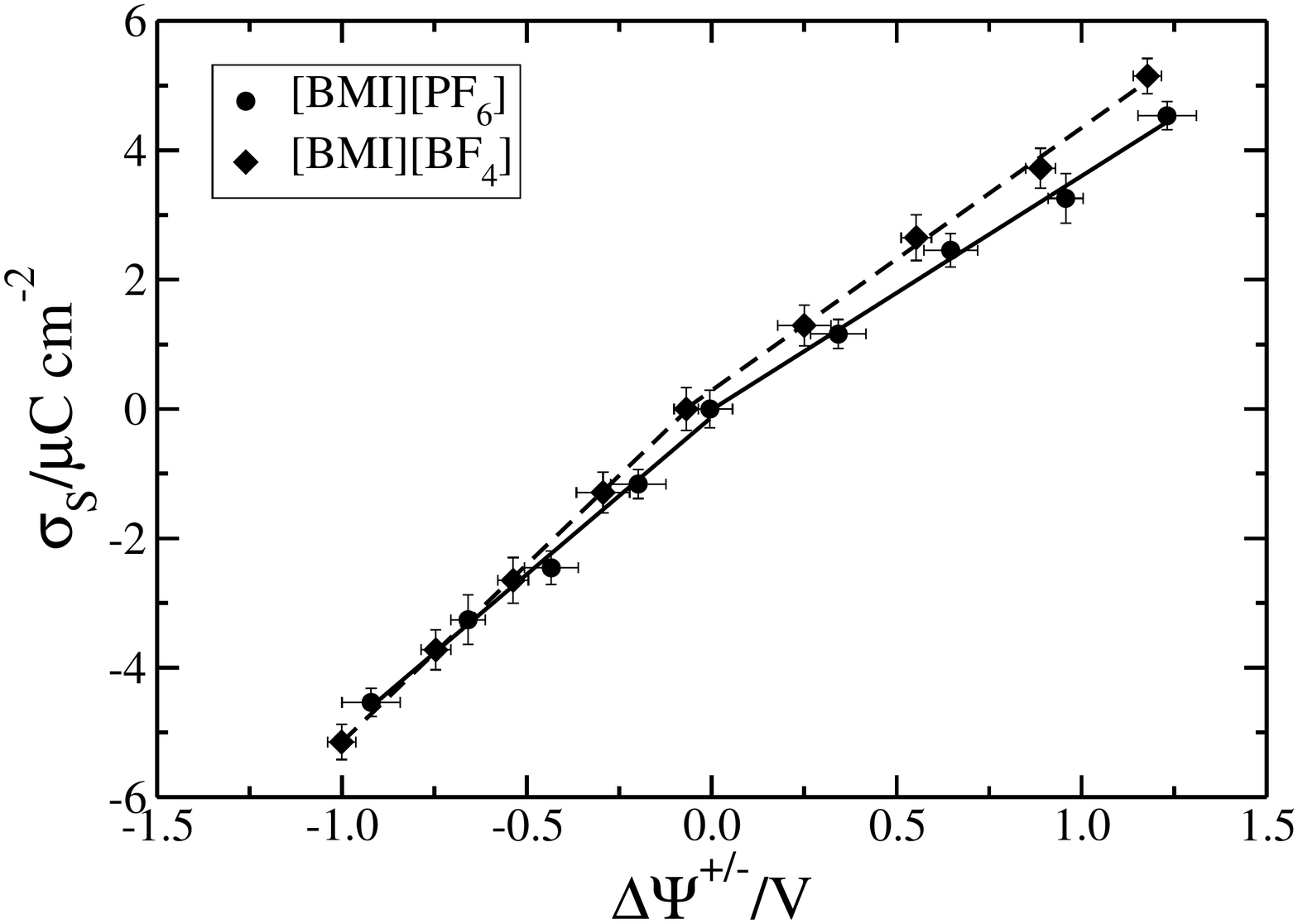}
\includegraphics[scale=0.27]{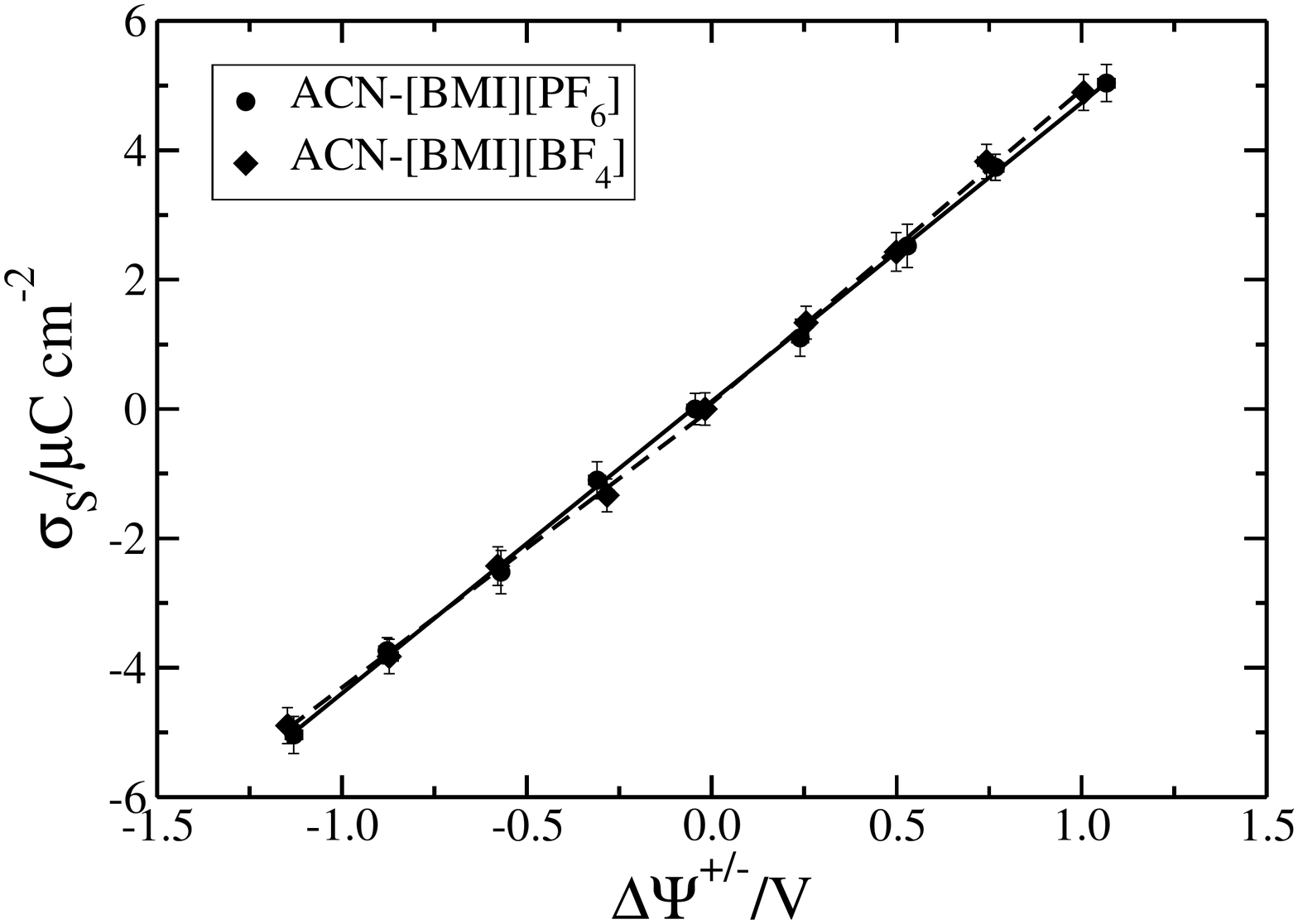}\\
\caption{Surface charge ($\sigma_{\rm S}$) with respect to the potential drop across the interface ($\Delta\Psi^{\pm} = \Psi^{\pm} - \Psi_{\rm bulk}$) for the electrolytes examined in this work. Error bars for the estimation of the two quantities are indicated on the graph. As for the charge distribution curves, the addition of a solvent results in a superimposition of the curves. Linear trends for positive and negative electrodes are visible for all electrolytes. In the case of electrolyte solutions, the points can be fitted by a single linear curve for all the potentials explored.}
\label{capa}
\end{figure*}

The presence of solvent has an effect on the molecular densities at the interface, on the polarization of the electrodes and consequently on the capacitive properties of the system. The differential capacitance of each interface depends on the surface charge on the electrode and on the potential drop across the interface. The surface charge is taken as the average total charge on the electrode divided by the surface area of one graphene layer. The potential drop is extracted from the electrostatic potential profile which is a function of the charge density and is described by Poisson's equation:
\begin{equation}
\Psi(z) = \Psi_q(z_0) - \frac{1}{\varepsilon_0}\int_{z_0}^zdz'\int_{-\infty}^{z'}dz''\rho_q(z''),
\end{equation} 
where $z_0$ is a reference point inside the left-hand electrode and thus, $\Psi_q(z_0)$~=~$\Psi^+$, $\varepsilon_0$ is the vacuum permittivity and $\rho_q(z)$ is the charge density. The two potential drops, depending on the considered electrode, are defined as follows~\cite{Pounds09,Tazi10,Merlet11}:
\begin{equation}
\Delta\Psi^{\pm} = \Psi^{\pm} - \Psi_{\rm bulk}, 
\end{equation}
and the differential capacitances, for the positive and negative electrodes, result from the differentiation of the surface charge with respect to these potential drops:
\begin{equation}
\rm C^{\pm} = \frac{\partial\sigma_{\rm S}}{\partial\Delta\Psi^{\pm}}.
\end{equation}
Figure~\ref{capa} gives the surface charge variations as a function of the potential drops for the studied electrolytes. All the functions plotted show linear trends over the range of potentials sampled and the differential capacitances are calculated as the slopes of these functions and gathered in table~\ref{capa_values}. We note that the capacitances were all evaluated separately for negative and positive electrodes but in the case of electrolyte solutions, in the light of the errors made in the estimation of the average charges and potential drops (see figure~\ref{capa}), it would be possible to fit the entire curve by a single linear function. 

\begin{table*}
\begin{center}
\begin{tabular}{|c|c|c|}
\hline
Electrolyte & C$^+$ ($\mu$F.cm$^{-2})$ & C$^-$ ($\mu$F.cm$^{-2})$ \\
\hline
[BMI][PF$_6$] & 3.9 ($\pm$~0.3) & 4.8 ($\pm$~0.5) \\
\hline
[BMI][BF$_4$] & 3.9 ($\pm$~0.3) & 5.5 ($\pm$~0.1) \\
\hline
ACN-[BMI][PF$_6$] & 4.6 ($\pm$~0.2) & 4.6 ($\pm$~0.2) \\
\hline
ACN-[BMI][BF$_4$] & 4.8 ($\pm$~0.2) & 4.3 ($\pm$~0.2) \\
\hline
\end{tabular}
\end{center}
\caption{Capacitance values obtained in the present work. Linear trends observed for the negative and positive electrodes lead to the estimation of two differential capacitances for each system. Error bars estimated for a confidence interval of 95~\% are given in parentheses.}
\label{capa_values}
\end{table*}
Looking at the surface charge versus potential drop plots and capacitance values, it clearly appears that the trend is the same as for the polarization of the electrodes: In the solutions, the behavior of the two salts become very similar on both the positive and negative sides, despite the strong asymmetry in size and shape between the ions (this is particularly true for [BMI][BF$_4$]). On the positive electrode side, the capacitance is increased when going from pure ILs to solvated ions, and the reverse is observed for the negative electrode side, leading to a mean value of around 4.6~$\mu$F.cm$^{-2}$ for both interfaces in the electrolyte solutions. 

This uniformization of the capacitive behaviors upon addition of a solvent is consistent with the molecular density profiles which are dominated by a small dependency on the electrode potential for ACN and with the charge distributions functions which are more gaussian shaped for electrolyte solutions compared to pure ILs. In the presence of ACN in the first adsorbed layer, specific adsorption effects due to the molecular details appear to be wiped off, even if at this concentration highly-diluted theories remain irrelevant.

We would like to point out here that we could expect a lower capacitance value for the diluted electrolytes compared to the pure ionic liquids as the screening efficiency decreases with a decrease of ionic concentration. Our results reveal that this not the case. It presumably reflects the fact that the solvent enables cations and anions to be more readily separated by the application of a potential difference so that the layer compensating the charge on the electrode is narrower consistently with the reduction of the overscreening in electrolyte solutions. 

The fact that the capacitance of each interface is nearly constant upon addition of a solvent was also observed in other molecular simulations. Feng \emph{et al.}~\cite{Feng11b} studied the interface between ACN-[BMI][BF$_4$] electrolytes and graphite with a mass fraction of ACN ranging between 0~\% and 50~\% (their highest mass fraction is slighlty smaller than our 63~\% mass fraction of ACN for the ACN-[BMI][BF$_4$] solution). They show that the capacitance of the system is nearly constant and comprised between 6.5~$\mu$F.cm$^{-2}$ and 7.0~$\mu$F.cm$^{-2}$. With a slightly different electrolyte, ACN-[EMI][BF$_4$], Shim \emph{et al.} also reach the conclusion that the capacitance depends only weakly on the presence of the solvent. Interestingly, they do not observe the uniformization of the negative and positive capacitances, and we observe quantitative differences with their results. Our simulation procedure differs a lot from theirs due to the use of a constant potential approach for the electrodes with coarse-grained electrolytes in our case, while these authors have used a constant charge method with all-atom force fields. This observation should be explored in future works. We note that in a completely distinct electrolyte consisting of [Li][PF$_6$] and solvent mixtures~\cite{Vatamanu12b,Xing12b}, an asymmetry between negative and positive electrodes was noticed but, in this case, the dissymmetry between the anion and the cation is much more important.

\section{Conclusions}
\label{Conclusion}

In this article we have examined the impact of the presence of non-ionic solvent on the structural, polarization and capacitive properties of the interfaces between planar graphitic electrodes and liquid electrolytes. The main effect of solvation on the structure of the interface is the reduction of the region where ionic/molecular layering is observed near the graphite electrodes. Although the density of ions at the surface is lower in the solutions, this reduction in layering appears to result in a smaller reduction in the capacitance with respect to the pure ILs than might be expected. The molecular density profiles also show that the ACN molecules are only weakly affected by the potential difference applied between the electrodes. The reorganization of the layers upon charging of the electrodes varies when going from pure ILs, in which a polarization of the ionic layers occur, to electrolyte solutions, where a mechanism based on exchange of ions between the different layers is at play. From the coordination numbers at the interface, we conclude that, for the electrolyte solutions, the coordination number of ACN molecules around ions is more affected by the interface than the coordination numbers between ions. 

The polarization of the electrode is highly influenced by the type of electrolyte present. Solvent-free electrolytes generates charge distributions functions with skewed shapes and are impacted by the nature of the ions. On the contrary, charge distributions curves for electrolyte solutions have shapes closer to gaussians and do not depend on the size of the anion. In pure ILs, the polarization of the electrode also leads to a larger dissymmetry between positive and negative applied potentials compared to electrolyte solutions because of the asymmetry of the ions. 

The effect of solvation extends to the capacitive properties of the system. When a solvent is present in the electrolyte, the size/asymmetry of the ions do not generate different capacitances and the two solutions have similar properties. Moreover, the dissymmetry between positive and negative potentials is attenuated and a general linear trend is observed for the surface charge versus potential drop curves. 
 
This work raises conclusions about equilibrium interfaces between planar graphitic electrodes and free-solvent/electrolyte solutions which cannot be extended straightforwardly to porous electrodes systems. The effect of solvation in supercapacitors including porous electrodes will require further work in order to understand experimental results and design new electrolytes/electrodes. The impact of solvation on dynamic properties and charge/discharge processes should also be investigated to go further into the understanding of supercapacitors.  

\section*{Acknowledgements}

We acknowledge the support of the French Agence Nationale de la Recherche (ANR) under grant ANR-2010-BLAN-0933-02 (`Modeling the Ion Adsorption in Carbon Micropores'). We are grateful for the computing resources on JADE (CINES, French National HPC) obtained through the project x2012096728. This work made use of the facilities of HECToR, the UK's national high-performance computing service, which is provided by UoE HPCx Ltd at the University of Edinburgh, Cray Inc and NAG Ltd, and funded by the Office of Science and Technology through EPSRC's High End Computing Programme.






\newpage

\newpage

\newpage

\newpage

\newpage

\newpage

\newpage

\newpage

\end{document}